\documentclass[12pt,eqno,epsf]{article}
\usepackage{amsmath,amssymb,graphicx}
\numberwithin{equation}{section}

\def\ignore#1{{}}

\newcounter{sxn}

\newcounter{axn}

\date{}

\newdimen\mybaselineskip
\renewcommand{\thefootnote}{\arabic{footnote}}

\newcommand{\beeq}{\begin{equation}}
\newcommand{\eneq}{\end{equation}}
\newcommand{\beqn}{\begin{eqnarray}}
\newcommand{\eeqn}{\end{eqnarray}}

\newcommand{\alp}{\alpha}
\newcommand{\bt}{\beta}
\newcommand{\gm}{\gamma}

\newcommand{\dlt}{\delta}

\newcommand{\tht}{\theta}

\newcommand{\vth}{\vartheta}

\newcommand{\lmd}{\lambda}
\newcommand{\Lmd}{\Lambda}
\newcommand{\sgm}{\sigma}
\newcommand{\Sgm}{\Sigma}
\newcommand{\Ups}{\Upsilon}
\newcommand{\vph}{\varphi}
\newcommand{\omg}{\omega}
\newcommand{\Omg}{\Omega}
\newcommand{\dalp}{\dot{\alpha}}

\newcommand{\be}{\begin{equation}}
\newcommand{\ee}{\end{equation}}
\newcommand{\bea}{\begin{eqnarray}}
\newcommand{\eea}{\end{eqnarray}}
\newcommand{\eql}{\!\!\!&=\!\!\!&}

\newcommand{\defa}{\!\!\!&\equiv\!\!\!&}

\newcommand{\toa}{\!\!\!&\to\!\!\!&}

\newcommand{\tl}[1]{\tilde{#1}}
\newcommand{\bdm}[1]{{\mbox{\boldmath $#1$}}}

\newcommand{\diag}{{\rm diag}}
\newcommand{\der}{\partial}
\newcommand{\dr}{\!\!d}
\newcommand{\Hc}{{\rm h.c.}}
\newcommand{\ie}{{i.e.}}
\newcommand{\id}{\mbox{\boldmath $1$}}

\newcommand{\udl}[1]{\underline{#1}}

\newcommand{\vev}[1]{\langle #1 \rangle}

\newcommand{\brkt}[1]{\left( #1 \right)}
\newcommand{\brc}[1]{\left\{ #1 \right\}}
\newcommand{\sbk}[1]{\left[ #1 \right]}
\newcommand{\abs}[1]{\left| #1 \right|}

\renewcommand{\Re}{{\rm Re}\,}
\renewcommand{\Im}{{\rm Im}\,}

\newcommand{\cB}{{\cal B}}
\newcommand{\cC}{{\cal C}}

\newcommand{\cL}{{\cal L}}

\newcommand{\cN}{{\cal N}}
\newcommand{\cO}{{\cal O}}
\newcommand{\cP}{{\cal P}}

\newcommand{\cS}{{\cal S}}

\newcommand{\cV}{{\cal V}}

\newcommand{\cW}{{\cal W}}
\newcommand{\cX}{{\cal X}}
\newcommand{\cY}{{\cal Y}}
 
\newcommand{\bE}{{\mathbb E}}

\newcommand{\bH}{{\mathbb H}}

\newcommand{\bT}{{\mathbb T}}
\newcommand{\bV}{{\mathbb V}}
\newcommand{\rmE}{{\rm E}}
\newcommand{\rmT}{{\rm T}}
\newcommand{\SE}{S_{\rmE}}
\newcommand{\VE}{V_{\rmE}}
\newcommand{\derE}{\der_{\rmE}}
\newcommand{\hc}{h_{\rm c}}

\newcommand{\suU}{\mbox{SU(2)}_{\mbox{\scriptsize\bf U}}}

\begin{document}
\thispagestyle{empty}

\baselineskip=12pt

%{\small \noindent \mydate \hfill }

\begin{flushright}
KEK-TH-2063 \\
WU-HEP-18-8
\end{flushright}

\baselineskip=25pt plus 1pt minus 1pt

\vskip 1.5cm

\begin{center}
{\LARGE\bf $\bdm{\cN=1}$ superfield description} \\ 
{\LARGE\bf of BPS solutions in 6D gauged} \\
{\LARGE\bf SUGRA with 3-branes}

\vspace{1.0cm}
\baselineskip=20pt plus 1pt minus 1pt

\normalsize

{\large\bf Hiroyuki Abe,}${}^1\!${\def\thefootnote{\fnsymbol{footnote}}
\footnote[1]{E-mail address: abe@waseda.jp}}
{\large\bf Shuntaro Aoki}${}^1\!${\def\thefootnote{\fnsymbol{footnote}}
\footnote[2]{E-mail address: shun-soccer@akane.waseda.jp}}
{\large\bf Sosuke Imai}${}^1\!${\def\thefootnote{\fnsymbol{footnote}}
\footnote[3]{E-mail address: s.i.sosuke@akane.waseda.jp}} \\
{\large\bf and Yutaka Sakamura}${}^{2,3}\!${\def\thefootnote{\fnsymbol{footnote}}
\footnote[4]{E-mail address: sakamura@post.kek.jp}}

\vskip 1.0em

${}^1${\small\it Department of Physics, Waseda University, \\ 
3-4-1 Ookubo, Shinjuku-ku, Tokyo 169-8555, Japan}

\vskip 1.0em

${}^2${\small\it KEK Theory Center, Institute of Particle and Nuclear Studies, 
KEK, \\ 1-1 Oho, Tsukuba, Ibaraki 305-0801, Japan} \\ \vspace{1mm}
${}^3${\small\it Department of Particles and Nuclear Physics, \\
SOKENDAI (The Graduate University for Advanced Studies), \\
1-1 Oho, Tsukuba, Ibaraki 305-0801, Japan}

\end{center}

\vskip 1.0cm
\baselineskip=20pt plus 1pt minus 1pt

\begin{abstract}
We provide $\cN=1$ superfield description of BPS backgrounds 
in six-dimensional supergravity (6D SUGRA) with 3-branes, 
which is compactified on a two-dimensional space. 
The brane terms induce the localized fluxes. 
%It is obtained by solving the background equations derived from 
%the superfield equations of motion. 
We find a useful gauge in which the background equations become significantly simple. 
This is not the Wess-Zumino gauge, and 
the relation to the usual component-field expression of 6D SUGRA is not straightforward.  
One of the equations reduces to the Liouville equation. 
By moving to the Wess-Zumino gauge, we check that our expressions 
reproduce the known results of the previous works, 
which are expressed in the component fields. 
Our results help us develop the systematic derivation 
of four-dimensional effective theories that keeps the $\cN=1$ SUSY structure. 
\end{abstract}

%%%%%%%%%%%%%%%%%%%%%%%%
\newpage

\section{Introduction}
The $\cN=1$ superfield description of higher-dimensional supersymmetric (SUSY) 
theories~\cite{Marcus:1983wb}-\cite{Abe:2004ar} 
is useful in various aspects.\footnote{
``$\cN=1$'' denotes SUSY with four supercharges in this paper. 
} 
It makes the expression of the action compact. 
Especially, when we consider a system with lower-dimensional branes whose dimensions 
are not less than four, we can express the action keeping the common $\cN=1$ SUSY 
manifest. 
It describes the bulk-brane interactions in a transparent manner. 
Such a description also makes it possible to derive four-dimensional (4D) effective action 
directly from the original higher-dimensional theory. 
Besides, the $\cN=1$ superfield formalism is familiar to many researchers, 
and easy to handle. 

When the extra dimensions are compactified on some manifold or orbifold, 
the moduli fields appear in 4D effective theories. 
In order to deal with such moduli and discuss the moduli stabilization, 
we have to work in the context of supergravity (SUGRA). 
In this paper, we consider six-dimensional (6D) SUGRA 
compactified on two-dimensional compact spaces, such as a sphere or torus. 
It is known that a tensor multiplet need to be introduced 
for the Lagrangian description of 6D SUGRA~\cite{Bergshoeff:1985mz,Coomans:2011ih}. 
We have derived the $\cN=1$ superfield description of the couplings 
between the tensor and vector multiplets in Ref.~\cite{Abe:2015bqa}, 
and extended the result to SUGRA by inserting the superfields that contain 
the fields in the Weyl multiplets in Refs.~\cite{Abe:2015yya,Abe:2017pvw}. 
The $\cN=1$ superfield action is also a good starting point to derive 4D effective theory 
keeping the $\cN=1$ SUSY structure, 
just we did in five-dimensional SUGRA~\cite{Abe:2006eg}-\cite{Sakamura:2014aja}. 

We can introduce brane terms localized in the extra dimensions. 
Here we add brane terms to the bulk action, 
which lead to the localized magnetic fluxes. 
They also induce the tensions of the branes, and affect the geometry 
of the compact space~\cite{Carroll:2003db,Navarro:2003vw,Navarro:2003bf}. 

From the superfield action, we can derive the superfield equations of motion (EOMs) straightforwardly. 
These equations become much simpler for 
the Bogomol'nyi-Prasad-Sommerfield (BPS) background that preserves $\cN=1$ SUSY. 
By solving those BPS equations, we obtain the BPS background field configurations. 
Such backgrounds have been investigated 
in the previous works, mainly for the sphere compactification~\cite{Carroll:2003db,Navarro:2003vw,Navarro:2003bf}. 
The backgrounds in these works are described in terms of the component fields in 6D SUGRA. 
In this paper, we express the BPS backgrounds in the $\cN=1$ superfield language. 
As we have shown in our previous work~\cite{Abe:2017pvw}, 
the relation between the superfields and the usual component fields of 6D SUGRA 
is not simple. 
Thus expressing the known background configurations in terms of the superfields 
is a nontrivial task. 
This helps us develop a systematic derivation of 4D effective theories 
that keeps the $\cN=1$ SUSY manifest, and enables us to treat the compactifications 
with different topologies on equal footing. 

The paper is organized as follows. 
In the next section, we briefly review our previous work~\cite{Abe:2017pvw}, 
and provide the bulk and the brane actions in terms of $\cN=1$ superfields. 
Sect.~\ref{BPS_bgd} is the main part of this paper. 
We derive the BPS background equations from the superfield action. 
After choosing an appropriate gauge, we solve the equations 
and check that the known results are reproduced by the solutions 
in our superfield approach. 
In Sects.~\ref{sphere_cmp} and \ref{torus_cmp}, we summarize the results 
in the previous works in our notation 
for the sphere and torus compactifications, respectively. 
Sect.~\ref{summary} is devoted to the summary. 
In Appendix~\ref{comp_fields}, we explicitly show some of the components of the superfields 
in terms of the usual component fields of 6D SUGRA. 
In Appendix~\ref{bgd_eq:U^m}, we list some of the EOMs that are not shown 
in the text because of their lengthy expressions. 
We provide a comment on the quantization condition of the total flux in Appendix~\ref{FQ:sphere}, 
and the definitions of the Weierstrass elliptic functions in Appendix~\ref{def:Welliptic}.

\section{$\bdm{\cN=1}$ superfield description of 6D SUGRA action}
In this section, we provide a brief review of our previous results in Ref.~\cite{Abe:2017pvw}. 
The 6D spacetime indices~$M,N,\cdots = 0,1,2,\cdots,5$ are divided into 
the 4D part~$\mu,\nu,\cdots=0,1,2,3$ and the extra-dimensional part~$m,n,\cdots=4,5$. 
The corresponding local Lorentz indices are denoted by the underbarred ones. 
We assume that the 4D part of the spacetime has the flat background geometry,\footnote{
In 6D SUGRA compactified on a 2D compact space, 
the 4D flat spacetime is a unique maximally symmetric solution~\cite{Gibbons:2003di,Lee:2005az}. 
}
and follow the notation of Ref.~\cite{Wess:1992cp} for the 2-component spinors.

\subsection{$\bdm{N=1}$ decomposition of 6D supermultiplets} \label{superfields}
The field content of 6D SUGRA consists of the Weyl multiplet~$\bE$, 
the hypermultiplets~$\bH^A$ ($A=1,2,\cdots,n_{\rm H}$), 
the vector multiplets~$\bV^I$ ($I=1,2,\cdots,n_{\rm V}$), 
and the tensor multiplet~$\bT$.\footnote{
We focus on the case of a single tensor multiplet 
because the theory cannot be described by the Lagrangian in the other cases. 
Besides, the anomaly cancellation conditions also constrain $n_{\rm H}$ and $n_{\rm V}$ 
and the gauge group~\cite{RandjbarDaemi:1985wc,Green:1984bx,Kumar:2010ru}. 
In this paper, we do not take account of such constraints, and assume that the gauge groups 
are Abelian, for simplicity. 
} 

Each 6D supermultiplet can be decomposed into $\cN=1$ superfields as follows. 
\begin{description}
\item[Weyl multiplet~$\bE$]
\bea
 U^\mu, \; U^4, \; U^5, \; \VE &:& \mbox{Real superfields} \nonumber\\
 \SE &:& \mbox{Chiral superfield} \nonumber\\
 \Psi_4^\alp, \; \Psi_5^\alp &:& \mbox{Spinor superfields}
 \label{sf:Weyl}
\eea
The superfields~$\VE$ and one of $\Psi_4^\alp$ and $\Psi_5^\alp$ are 
dependent fields, as will be mentioned in Sect.~\ref{tensor_cstrt}.

\item[Hypermultiplet~$\bH^A$]
\be
 H^{2A-1}, \; H^{2A} \;\; : \;\; \mbox{Chiral superfields}
 \label{sf:hyper}
\ee
The hypermultiplets are divided into the compensator multiplets~$A=1,2,\cdots,n_{\rm comp}$ 
and the physical ones~$A=n_{\rm comp}+1,\cdots,n_{\rm H}$.

\item[Vector multiplet~$\bV^I$]
\bea
 V^I &:& \mbox{Real superfield} \nonumber\\
 \Sgm^I &:& \mbox{Chiral superfield}
\eea

\item[Tensor multiplet~$\bT$]
\bea
 \Ups_{\rm T\alp} &:& \mbox{Chiral spinor superfield} \nonumber\\
 V_{\rm T4}, \; V_{\rm T5} &:& \mbox{Real superfields} \nonumber\\
 \Sgm_{\rm T} &:& \mbox{Chiral superfield}
\eea
\end{description}
The correspondence between these superfields and the component fields of 6D SUGRA 
is summarized in Appendix~\ref{comp_fields}. 
The Weyl weights of the superfields are listed in Table~\ref{Weyl_weight}.\footnote{
We should note that $V^I$ in (\ref{comp:V^I}) 
(and $U^\mu$, $U^m$ in (\ref{comp:U^mu}), $V_{\rmT m}$ in (\ref{comp:Ups_T}))
are in the Wess-Zumino gauge. 
This indicates that we need to choose the Wess-Zumino gauge 
in order to see the correspondence to the component-field expression of 6D SUGRA. 
\label{sf-comp:corr}
} 
%%%%%%%%%%%%%%%%%%%%%%%%%%%%% Table %%%%%%%%%%%%%%%%%%%%%%%%%%%%%%%%%%%%%%%%
\begin{table}[t]
\begin{center}
\begin{tabular}{|c|c|c|c|c|c|c|c|c|c|c||c|c|c|c|} \hline
\multicolumn{5}{|c|}{$\bE$} & $\bH^A$ & \multicolumn{2}{|c|}{$\bV^I$} & 
\multicolumn{3}{|c||}{$\bT$} & \multicolumn{4}{|c|}{field strength} \\\hline
$U^\mu$ & $U^m$ & $\Psi_m^\alp$ & $\SE$ & $V_{\rm E}$ & 
$H^{\bar{A}}$ & $V^I$ & $\Sgm^I$ & $\Ups_{\rmT\alp}$ & $V_{\rmT m}$ & $\Sgm_{\rm T}$ 
& $\cW^I_\alp$ & $\cX_{\rmT}$ & $\cY_{\rmT\alp}$ & $\cV_{\rmT}$ \\\hline
0 & 0 & $-3/2$ & 0 & $-2$ & $3/2$ & 0 & 0 & $3/2$ & 0 & 0 & $3/2$ & 2 & $3/2$ & 0 \\\hline
\end{tabular}
\end{center}
\caption{The Weyl weights of the $\cN=1$ superfields. 
The index~$\bar{A}$ runs from 1 to $2n_{\rm H}$. }
\label{Weyl_weight}
\end{table}
%%%%%%%%%%%%%%%%%%%%%%%%%%%%%%%%%%%%%%%%%%%%%%%%%%%%%%%%%%%%%%%%%%%%%%%%%%%

Among the above superfields, $U^\mu$ corresponds to the 4D part of the Weyl multiplet, 
and will be dropped in the following expressions 
because they are irrelevant to the background equations.\footnote{
The $U^\mu$-dependence of the action 
can be easily recovered by using the result of Ref.~\cite{Sakamura:2011df}, 
in which the linearized 4D SUGRA is discussed. 
}

The superfields~$U^m$ contain $e_{\udl{\mu}}^{\;\;m}$, 
and are used to covariantize the spinor derivatives~$D_\alp$ and $\bar{D}_{\dalp}$. 
Define the operator~$\cP_U$ that shifts $x^m$ by $iU^m$. 
\be
 \cP_U :\; x^m \to x^m+iU^m(x,\tht,\bar{\tht}). 
\ee
Then, the covariant derivatives are defined as
\be
 D^{\cP}_\alp \equiv \bar{\cP}_UD_\alp\bar{\cP}_U^{-1}, \;\;\;\;\;
 \bar{D}^{\cP}_{\dalp} \equiv \cP_U\bar{D}_{\dalp}\cP_U^{-1}. 
\ee

The superfields~$\Psi_m^\alp$ contain $e_m^{\;\;\udl{\mu}}$, 
and are used to covariantize $\der_m$ as~\footnote{
We need not discriminate the flat 4D index~$\udl{\mu}$ and the curved one~$\mu$ 
at the linearized order since the 4D part of the background spacetime is assumed to be flat
($\vev{e_\nu^{\;\;\udl{\mu}}}=\dlt_\nu^{\;\;\mu}$).  
}
\be
 \nabla_m \equiv \der_m-\brkt{\frac{1}{4}\bar{D}^2\Psi_m^\alp D_\alp
 -i\sgm_{\alp\dalp}^{\udl{\mu}}\bar{D}^{\dalp}\Psi_m^\alp\der_\mu
 +\frac{w}{12}\bar{D}^2D^\alp\Psi_{m\alp}}+\cO(\Psi^2,\Psi U^n), 
\ee
in the chiral superspace, and 
\bea
 \nabla_m^\cP \defa \cP_U\der_m\cP_U^{-1}-\brkt{\frac{1}{4}\bar{D}^2\Psi_m^\alp D_\alp
 +\frac{1}{2}\bar{D}^{\dalp}\Psi_m^\alp\bar{D}_{\dalp}D_\alp
 +\frac{w+n}{24}\bar{D}^2D^\alp\Psi_{m\alp}} \nonumber\\
 &&-\brkt{\frac{1}{4}D^2\bar{\Psi}_{m\dalp}\bar{D}^{\dalp}
 +\frac{1}{2}D^\alp\bar{\Psi}_m^{\dalp}D_\alp\bar{D}_{\dalp}
 +\frac{w-n}{24}D^2\bar{D}_{\dalp}\bar{\Psi}_m^{\dalp}}+\cO(\Psi^2,\Psi U^n), 
\eea
in the full superspace. 
Here, $w$ and $n$ denote the Weyl and chiral weights, respectively.

The (super) gauge transformations are given by
\be
 \dlt_\Lmd V^I = \Lmd^I+\bar{\Lmd}^I, \;\;\;\;\;
 \dlt_\Lmd\Sgm^I = \nabla_{\rm E}\Lmd^I,  \label{GF_trf}
\ee
where the transformation parameters~$\Lmd^I$ are chiral superfields, and 
\be
 \nabla_{\rm E} \equiv \frac{1}{\SE}\nabla_4-\SE\nabla_5. 
\ee
The gauge-invariant field strength superfields are given by
\be
 \cW_\alp^I \equiv -\frac{1}{4}\brkt{\bar{D}^\cP}^2D^\cP_\alp V^I
 +\cO(U^m\Sgm). 
\ee

The SUSY extension of the tensor gauge transformation:~$B_{MN}\to B_{MN}+\der_M\lmd_N-\der_N\lmd_M$ 
($\lmd_M$: real parameter) is expressed as 
\bea
 \dlt_{\rm G}V_{\rm T4} \eql -\der_4V_{\rm G}+\Re\brkt{\SE\Sgm_{\rm G}}, \;\;\;\;\;
 \dlt_{\rm G}V_{\rm T5} = -\der_5V_{\rm G}+\Re\brkt{\frac{\Sgm_{\rm G}}{\SE}}, \nonumber\\
 \dlt_{\rm G}\Ups_{\rm T\alp} \eql -\frac{1}{4}\bar{D}^2D_\alp V_{\rm G}, \nonumber\\
 \dlt_{\rm G}\Sgm_{\rmT} \eql -\frac{1}{2}\der_4\brkt{\frac{\Sgm_{\rm G}}{\SE}}
 +\frac{1}{2}\der_5\brkt{\SE\Sgm_{\rm G}}, 
\eea
up to $U^m$- or $\Psi_m^\alp$-dependent terms. 
The transformation parameters~$V_{\rm G}$ and $\Sgm_{\rm G}$ are real 
and chiral superfields respectively, which form a 6D vector multiplet~$\bV_{\rm G}$. 
The field strength superfields invariant under this transformation are~\footnote{
The $U^m$- and $\Psi_m^\alp$-dependences are determined by the (SUSY extension of) diffeomorphism 
covariance~\cite{Abe:2017pvw}. 
} 
\bea
 \cX_{\rm T} \defa \frac{1}{2}\Im\brkt{D^{\cP\alp}\hat{\Ups}_{\rm T\alp}}, \nonumber\\
 \cY_{\rm T} \defa \frac{1}{2\SE}\cW_{\rmT 4\alp}+\frac{\SE}{2}\cW_{\rmT 5\alp}
 +\frac{1}{2}\SE\cO_{\rm E}\Ups_{\rmT\alp}, \nonumber\\
 \cV_{\rmT} \defa \Re\brkt{\nabla_4^\cP V_{\rmT 5}-\nabla_5^\cP V_{\rmT 4}
 +2J_\cP\hat{\Sgm}_{\rmT}},
\eea
where
\bea
 \hat{\Ups}_{\rmT\alp} \defa \cP_U\Ups_{\rmT\alp}, \nonumber\\
 \cW_{\rmT m\alp} \defa -\frac{1}{4}\brkt{\bar{D}^\cP}^2D_\alp^\cP V_{\rmT m}
 +\cO(U^n\Sgm_{\rmT}), \nonumber\\
 \cO_{\rm E} \defa \frac{1}{\SE^2}\nabla_4+\nabla_5, \;\;\;\;\;
 \hat{\Sgm}_{\rm T} \equiv \cP_U\Sgm_{\rm T}, \nonumber\\
 J_\cP \defa 1+i\der_m U^m-\der_4 U^4\der_5 U^5+\der_4 U^5\der_5 U^4. 
\eea
Note that $J_\cP$ is the Jacobian for the shift by $\cP_U$.

\subsection{Constraints on tensor multiplets} \label{tensor_cstrt}
The tensor multiplet~$(\Ups_{\rmT\alp},V_{\rmT m})$ is subject to the following the constraints, 
which reduce to the self-dual condition in the global SUSY limit. 
\bea
 &&\cX_{\rm T}\VE = \cV_{\rm T}, \nonumber\\
 &&\frac{1}{\SE}\cW_{\rmT 4\alp}-\SE\cW_{\rmT 5\alp}+\nabla_{\rm E}\Ups_{\rmT\alp} = 0. 
\eea
From the first constraint, the ``volume modulus'' superfield~$\VE$ is expressed 
in terms of the tensor field strengths~$\cX_{\rmT}$ and $\cV_{\rmT}$. 
Since $\nabla_{\rm E}$ depends on $\Psi_m^\alp$, 
the second constraint indicates that either $\Psi_4^\alp$ or $\Psi_5^\alp$ 
are dependent field, and can be expressed in terms of the other superfields.

\subsection{Invariant action} \label{inv_action}
We will omit $U^\mu$ and $\Psi_m^\alp$ in the following 
because they are irrelevant to the discussions in the next sections.

\subsubsection{Bulk action}
The 6D SUGRA action is expressed in terms of the $\cN=1$ superfields as
\bea
 S_{\rm bulk} \eql \int d^6x\;\brkt{\cL_{\rm H}+\cL_{\rm VT}}, \nonumber\\
 \cL_{\rm H} \eql -2\int\dr^4\tht\;
 \abs{J_\cP}\brkt{\frac{\cV_{\rmT}R_{\rm E}^-}{\cX_{\rmT}}}^{1/2}
 \brkt{\hat{H}_{\rm odd}^\dagger\tl{d}e^{V}\hat{H}_{\rm odd}
 +\hat{H}_{\rm even}^\dagger\tl{d}e^{-V}\hat{H}_{\rm even}}, \nonumber\\
 &&+\sbk{\int\dr^2\tht\;\brc{H_{\rm odd}^t\tl{d}\brkt{\derE-\Sgm}H_{\rm even}
 -H_{\rm even}^t\tl{d}\brkt{\derE+\Sgm}H_{\rm odd}}+\Hc}, \nonumber\\
 \cL_{\rm VT} \eql \int\dr^4\tht\;f_{IJ}\left[\brc{
 -2J_\cP\hat{\Sgm}^ID^{\cP\alp}V^J\hat{\cY}_{\rmT\alp}
 +\frac{J_\cP}{2}\brkt{\derE^\cP V^I D^{\cP\alp}V^J-\derE^\cP D^{\cP\alp}V^IV^J}\hat{\cY}_{\rmT\alp}+\Hc}
 \right. \nonumber\\
 &&\hspace{20mm}
 +\cV_{\rmT}\brkt{D^{\cP\alp}V^I\hat{\cW}_\alp^J
 +\frac{1}{2}V^I D^{\cP\alp}\hat{\cW}_\alp^J+\Hc} \nonumber\\
 &&\hspace{20mm}
 +\frac{\cX_{\rmT}}{R_{\rm E}^-}\left\{4\brkt{\derE^\cP V^I-\hat{\Sgm}^I}^\dagger
 \brkt{\derE^\cP V^J-\hat{\Sgm}^J}-2\brkt{\derE^\cP V^I}^\dagger\derE^\cP V^J \right. \nonumber\\
 &&\hspace{33mm}
 +\brkt{2J_\cP R_{\rm E}^+\hat{\Sgm}^I\hat{\Sgm}^J+\Hc}\bigg\}\bigg], 
\eea
where $H_{\rm odd}\equiv (H^1,H^3,\cdots,H^{2n_{\rm H}-1})^t$ 
$H_{\rm even}\equiv (H^2,H^4,\cdots,H^{2n_{\rm H}})^t$, 
$\hat{\Phi}\equiv \cP_U\Phi$ for a chiral superfield~$\Phi$, 
$\derE\equiv \SE^{-1}\der_4-\SE\der_5$, $\derE^\cP\equiv\cP_U\derE\cP_U^{-1}$, and 
\bea
 R_{\rm E}^- \defa \frac{1}{2i}\brkt{J_S^{(2)}\frac{\bar{\hat{S}}_{\rm E}}{\hat{S}_{\rm E}}
 -J_S^{(1)}\frac{\hat{S}_{\rm E}}{\bar{\hat{S}}_{\rm E}}}, \nonumber\\
 R_{\rm E}^+ \defa \frac{1}{2}\brkt{J_S^{(2)}\frac{\bar{\hat{S}}_{\rm E}}{\hat{S}_{\rm E}}
 +J_S^{(1)}\frac{\hat{S}_{\rm E}}{\bar{\hat{S}}_{\rm E}}}, \nonumber\\
 J_S^{(1)} \defa 1+i\brkt{\der_4 U^4-\der_5 U^5}-2i\bar{\hat{S}}_{\rm E}^2\der_5 U^4+\cO(U^2), \nonumber\\
 J_S^{(2)} \defa 1-i\brkt{\der_4 U^4-\der_5 U^5}-\frac{2i}{\bar{\hat{S}}_{\rm E}^2}\der_4 U^5+\cO(U^2). 
\eea
The $n_{\rm H}\times n_{\rm H}$ constant matrix~$\tl{d}$ is the metric of the hyperscalar space that discriminates 
the compensator multiplets from the physical ones, and can be chosen as
$\tl{d}=\diag(\id_{n_{\rm comp}},-\id_{n_{\rm H}-n_{\rm comp}})$.  
The $n_{\rm V}\times n_{\rm V}$ constant matrix~$f_{IJ}$ is real and symmetric. 
In the hyper-sector Lagrangian~$\cL_{\rm H}$, the vector multiplets are described 
in the matrix notation, 
\be
 V \equiv V^It_I, \;\;\;\;\;
 \Sgm \equiv \Sgm^It_I, 
\ee
where $t_I$ are the generators for the Abelian gauge group, \ie, the charge matrices. 
Their components are denoted as 
\be
 t_I = \begin{pmatrix} 2c_I & \\ & \ddots \end{pmatrix}, 
\ee
where $c_I$ are the compensator charges. 

The above action is invariant under the diffeomorphisms 
and the Lorentz transformations involving the extra dimensions, 
and the (super) gauge transformations~\cite{Abe:2017pvw}.

\subsubsection{Brane action}
We also introduce brane terms localized at $x^m=x_k^m$ ($k=1,\cdots,N$).\footnote{
We do not consider branes whose codimension is one, for simplicity. 
} 
Here we consider the case of single compensator, \ie, $n_{\rm comp}=1$. 
Since the bulk and the branes feel the same gravity, 
the chiral compensator superfields appearing in the brane action
should originate from the bulk compensator multiplet~$\bH^1=(H^1,H^2)$. 
We should note that $H^1$ and $H^2$ cannot mix with each other 
when $c_I\neq 0$ for some $I$ because they have opposite charges. 
Thus the brane compensators are either $H^1$ or $H^2$. 
For simplicity, we assume that all the brane compensators come from $H_{\rm even}^1=H^2$. 
Then we can introduce the following brane terms. 
\bea
 S_{\rm brane} \eql \int\dr^6x\;\cL_{\rm brane}, \nonumber\\
 \cL_{\rm brane} \eql -\int\dr^4\tht\;\sum_{k=1}^N C_k
 \brkt{\frac{\cX_{\rmT}R_{\rm E}^-}{\cV_{\rmT}}}^{1/4}
 \brkt{\bar{\hat{H}}_{\rm even}^1 e^{-2c_IV^I}\hat{H}_{\rm even}^1}^{1/2}\dlt^{(2)}(y-y_k),  \label{L_brane}
\eea
where $C_k$ are real constants, $\vec{y}\equiv (x^4,x^5)^t$ are the extra-dimensional coordinates, 
and $\vec{y}_k\equiv (x^4_k,x^5_k)^t$ are the brane positions. 
The powers in (\ref{L_brane}) are determined by the Weyl weight 
and by requiring that the extra-dimensional components of the sechsbein~$e_m^{\;\;\udl{n}}$ 
contained in the superfields are cancelled. 
(See (\ref{lowest:R_E}), (\ref{comp:hyper}), and (\ref{comp:cV_T}).)
The above terms represent the brane-localized Fayet-Iliopoulos (FI) terms, 
which lead to the brane tensions and the localized fluxes as we will see in the next section.

\section{BPS Background} \label{BPS_bgd}
\subsection{Background equations of motion}
By varying the action in the previous section with respect to the superfields, 
we obtain the superfield EOMs. 
The background field configuration can be found by solving them. 
Here, we focus on the background that preserves $\cN=1$ SUSY. 
Namely, the F- and D-terms of the superfields can be put to zero. 
Besides, since we are interested in the 4D-Lorentz-invariant background, 
all the fermionic components and the bosonic components with the Lorentz indices 
are assumed to have vanishing backgrounds. 
This means that 
$U^m=0$, and $\SE$, $V_{\rmT m}$, $\Sgm_{\rmT}$, $H_{\rm even}$, $H_{\rm odd}$, 
$V^I$, $\Sgm^I$ and $D^\alp\Ups_{\rmT\alp}$ are $x^\mu$- and $\tht$-independent.\footnote{
We do not choose the Wess-Zumino gauge for $V_{\rmT m}$ and $V^I$. 
So their lowest components can have non-vanishing backgrounds. 
} 
Thus the tensor field strengths can be expressed as
\be
 \cX_{\rmT} = \frac{1}{2}\Im\brkt{D^\alp\Ups_{\rmT\alp}}, \;\;\;\;\;
 D^\alp\cY_{\rmT\alp} = \frac{1}{2}\SE\cO_{\rm E}\brkt{D^\alp\Ups_{\rmT\alp}}. 
\ee
Then we have the following EOMs for the background. 
\begin{description}
\item[For $\bdm{\SE}$]
\be
 \frac{1}{\SE^2}\brkt{H_{\rm odd}^t\tl{d}\der_4 H_{\rm even}-H_{\rm even}^t\tl{d}\der_4H_{\rm odd}}
 +\brkt{H_{\rm odd}^t\tl{d}\der_5H_{\rm even}-H_{\rm even}^t\tl{d}\der_5H_{\rm odd}} = 0. 
 \label{EOM:S_E}
\ee

\item[For $\bdm{V_{\rmT 4}}$ and $\bdm{V_{\rmT 5}}$]
\be
 \der_4\brc{\brkt{\frac{R_{\rm E}}{\cX_{\rmT}\cV_\rmT}}^{1/2}L_{\rm H}}
 = \der_5\brc{\brkt{\frac{R_{\rm E}}{\cX_{\rmT}\cV_\rmT}}^{1/2}L_{\rm H}} = 0, 
\ee
where
\be
 L_{\rm H} \equiv H_{\rm odd}^\dagger\tl{d}e^VH_{\rm odd}
 +H_{\rm even}^\dagger\tl{d}e^{-V}H_{\rm even}.  \label{def:L_H}
\ee

\item[For $\bdm{H_{\rm even}}$]
\be
 \brkt{\derE-\frac{1}{2}\cO_{\rm E}\SE+\Sgm}H_{\rm odd} = 0. 
\ee

\item[For $\bdm{H_{\rm odd}}$]
\be
 \brkt{\derE-\frac{1}{2}\cO_{\rm E}\SE-\Sgm}H_{\rm even} = 0. 
\ee

\item[For $\bdm{V^I}$]
\bea
 &&-2\brkt{\frac{\cV_\rmT R_{\rm E}}{\cX_\rmT}}^{1/2}
 \brkt{H_{\rm odd}^\dagger\tl{d}e^{V}t_IH_{\rm odd}
 -H_{\rm even}^\dagger\tl{d}e^{-V}t_I H_{\rm even}}
 +2f_{IJ}\brkt{\Sgm^J D^\alp\cY_{\rmT\alp}+\Hc} \nonumber\\
 &&-\der_4\brc{\frac{2f_{IJ}\cX_\rmT}{R_{\rm E}\bar{S}_{\rm E}}
 \brkt{\derE V^J-2\Sgm^J}+\Hc}
 +\der_5\brc{\frac{2f_{IJ}\cX_\rmT\bar{S}_{\rm E}}{R_{\rm E}}
 \brkt{\derE V^J-2\Sgm^J}+\Hc} \nonumber\\
 &&+\sum_{k=1}^N c_IC_k\brkt{\frac{\cX_\rmT R_{\rm E}}{\cV_\rmT}}^{1/4}
 \abs{H_{\rm even}^1}e^{-c_JV^J}\dlt^{(2)}(y-y_k) = 0. 
\eea

\item[For $\bdm{\Sgm^I}$]
\be
 H_{\rm odd}^t\tl{d}t_IH_{\rm even} = 0. \label{EOM:Sgm}
\ee
\end{description}
The EOMs for $U^4$ and $U^5$ are shown in (\ref{EOM:U4}) and (\ref{EOM:U5}).

\subsection{Coordinate and gauge choices}
It is convenient to choose the coordinates of the extra dimensions such that~\footnote{
We can always move to this coordinate system by using the (super) diffeomorphism 
(\ie, the $\dlt_\Xi$-transformation in Ref.~\cite{Abe:2017pvw}). 
}
\be
 \vev{\SE} = e^{-\pi i/4} \equiv \eta. 
\ee
Then, we have 
\bea
 R_{\rm E}^- \eql 1, \;\;\;\;\;
 R_{\rm E}^+ = 0, \nonumber\\
 \derE \eql 2\bar{\eta}\der_{\bar{z}}, \;\;\;\;\;
 \cO_{\rm E} = 2i\der_z, 
\eea
where $z\equiv x^4+ix^5$. 

We can gauge away the background of $\Sgm^I$ by using the transformation~(\ref{GF_trf}). 
Then the background EOMs~(\ref{EOM:S_E})-(\ref{EOM:Sgm}) are rewritten as
\bea
 &&H_{\rm odd}^t\tl{d}\der_z H_{\rm even}
 -H_{\rm even}^t\tl{d}\der_z H_{\rm odd} = 0, \nonumber\\
 &&\der_z\brkt{\frac{L_{\rm H}}{\sqrt{\cX_\rmT\cV_\rmT}}} 
 = \der_{\bar{z}}\brkt{\frac{L_{\rm H}}{\sqrt{\cX_\rmT\cV_\rmT}}} = 0, \nonumber\\
 &&\der_{\bar{z}}H_{\rm even} = \der_{\bar{z}}H_{\rm odd} = 0, \nonumber\\
 &&-2\brkt{\frac{\cV_\rmT}{\cX_\rmT}}^{1/2}\brkt{H_{\rm odd}^\dagger\tl{d}e^{V}t_IH_{\rm odd}
 -H_{\rm even}^\dagger\tl{d}e^{-V}t_IH_{\rm even}}
 -\brc{8f_{IJ}\der_z\brkt{\cX_\rmT\der_{\bar{z}}V^J}+\Hc} \nonumber\\
 &&+\sum_k 2c_IC_k\brkt{\frac{\cX_\rmT}{\cV_\rmT}}^{1/4}\abs{H_{\rm even}^1}e^{-c_JV^J}
 \dlt^{(2)}(z-z_k) = 0, \nonumber\\
 &&H_{\rm odd}^t\tl{d}t_IH_{\rm even} = 0, 
 \label{bgd:EOM}
\eea
where $z_k\equiv x_k^4+ix_k^5$. 
We have used that $\dlt^{(2)}(y-y_k)=2\dlt^{(2)}(z-z_k)$. 
The second equations can be solved as 
\be
 \frac{L_{\rm H}}{\sqrt{\cX_\rmT\cV_\rmT}} \equiv b_{\rm H} = (\mbox{real constant}). 
 \label{def:b_H}
\ee
The EOMs for $U^m$ are now written as  
\bea
 0 \eql -\brkt{\frac{\cV_\rmT^{1/2}L_{\rm H}}{\cX_\rmT^{3/2}}
 +L_{\rm V}}\der_z\Re\brkt{D^\alp\Ups_{\rmT\alp}} 
 +2i\der_z\brc{\brkt{\frac{\cV_\rmT}{\cX_\rmT}}^{1/2}L_{\rm H}+\cX_\rmT L_{\rm V}} 
 \nonumber\\
 &&-8\brkt{\frac{\cV_\rmT}{\cX_\rmT}}^{1/2}\Im\brkt{
 H_{\rm even}^\dagger\tl{d}e^{-V}\der_zH_{\rm even}
 +H_{\rm odd}^\dagger\tl{d}e^V\der_z H_{\rm odd}} \nonumber\\
 &&-2f_{IJ}\brc{i\bar{\eta}\brkt{\der_{\bar{z}}V^I\der_z V^J-\der_z\der_{\bar{z}}V^IV^J}
 D^\alp\cY_{\rmT\alp}
 -i\eta\brkt{\der_zV^I\der_zV^J-\der_z^2V^IV^J}\bar{D}_{\dalp}\bar{\cY}_\rmT^{\dalp}} \nonumber\\
 &&-2f_{IJ}\left\{
 i\bar{\eta}\der_{\bar{z}}\brkt{\der_zV^IV^JD^\alp\cY_{\rmT\alp}}
 -i\eta\der_z\brkt{\der_z V^IV^J\bar{D}_{\dalp}\bar{\cY}_\rmT^{\dalp}} \right.\nonumber\\
 &&\hspace{15mm} \left.
 +8i\der_z\brkt{\cX_\rmT\der_zV^I\der_{\bar{z}}V^J}
 -8i\der_{\bar{z}}\brkt{\cX_\rmT\der_zV^I\der_zV^J} \right\}+(\mbox{brane terms}), 
 \label{bgd:EOM:U}
\eea
by combining (\ref{EOM:U4}) and (\ref{EOM:U5}). 
We have used (\ref{def:b_H}), 
and $L_{\rm V}$ defined by (\ref{def:L_V}) becomes
\be
 L_{\rm V} = 8f_{IJ}\der_z V^I\der_{\bar{z}}V^J. 
\ee

\subsection{Background solution}
For simplicity, we consider a case of $n_{\rm V}=1$, 
and omit the indices~$I$ and $J$ in the following. 
Besides, we focus on a case that only $H_{\rm even}^1=H^2$ has a non-vanishing background value 
among $H^{\bar{A}}$. 
Then, it must be a constant from (\ref{bgd:EOM}). 
\bea
 H_{\rm even}^1 \defa \hc = (\mbox{complex constant}), \nonumber\\
 H_{\rm even}^{a\neq 1} \eql H_{\rm odd}^b = 0. 
\eea
Thus, $L_{\rm H}$ in (\ref{def:L_H}) is expressed as
\be
 L_{\rm H} = \abs{\hc}^2e^{-2cV}. 
\ee 

Here we denote the bosonic component of $\Ups_{\rmT\alp}$ as
\be
 D^\alp\Ups_{\rmT\alp} = B+2i\sgm, 
\ee
where $B$ and $\sgm$ are real. 
Then, $\cX_\rmT$ and $D^\alp\cY_{\rmT\alp}$ are expressed as
\be
 \cX_\rmT = \sgm, \;\;\;\;\;
 D^\alp\cY_{\rmT\alp} = \bar{\eta}\der_z\brkt{B+2i\sgm}. 
\ee

Using these results, the background EOM for the vector superfield~$V$ in (\ref{bgd:EOM}) becomes  
\be
 \frac{4c\abs{\hc}^4}{b_{\rm H}\sgm}e^{-4cV}-16f\Re\der_z\brkt{\sgm\der_{\bar{z}}V}
 +\sum_k2cC_k\sqrt{b_{\rm H}\sgm}\dlt^{(2)}(z-z_k) = 0, \label{EOM:V}
\ee
and (\ref{bgd:EOM:U}) becomes
\bea
 &&\frac{i\abs{\hc}^4e^{-4cV}}{2b_{\rm H}\sgm^2}\der_z B
 -2if\der_z V\brc{-2\der_{\bar{z}}V\der_zB+\brkt{\der_zV+\der_{\bar{z}}V}\der_{\bar{z}}B
 +V\der_z\der_{\bar{z}}B}  \nonumber\\
 &&+\der_z\brkt{\frac{\abs{\hc}^4e^{-4cV}}{b_{\rm H}\sgm}}
 +8f\der_z V\Re\brkt{\der_z\sgm\der_{\bar{z}}V+2\sgm\der_z\der_{\bar{z}}V}
 +(\mbox{brane terms}) = 0. 
\eea
%This is further rewritten as 
%\bea
% &&i\brkt{\frac{\abs{\hc}^4e^{-4cV}}{2b_{\rm H}\sgm^2}+4f\der_zV\der_{\bar{z}}V}\der_zB
% -2if\der_zV\Re\brkt{2\der_zV\der_{\bar{z}}B+V\der_z\der_{\bar{z}}B} \nonumber\\
% &&-\der_zV\brc{\frac{4c\abs{\hc}^4e^{-4cV}}{b_{\rm H}\sgm}
% -16f\Re\der_z\brkt{\sgm\der_{\bar{z}}V}}-8f\der_zV\Re\brkt{\der_z\sgm\der_{\bar{z}}V} \nonumber\\
% &&+(\mbox{brane terms}) = 0. 
%\eea
Using (\ref{EOM:V}), the latter becomes~\footnote{
The ``brane terms'' here contain the terms proportional to $\der_zV\dlt^{(2)}(z-z_k)$. 
Since $\der_zV$ has a singularity at $z=z_k$ as we will see, 
these ``brane terms'' are regularization-dependent and we do not evaluate them 
in this paper. 
}
\bea
  &&i\brkt{\frac{\abs{\hc}^4e^{-4cV}}{2b_{\rm H}\sgm^2}+4f\der_zV\der_{\bar{z}}V}\der_zB
 -2if\der_zV\Re\brkt{2\der_zV\der_{\bar{z}}B+V\der_z\der_{\bar{z}}B} \nonumber\\
 &&-8f\der_zV\Re\brkt{\der_z\sgm\der_{\bar{z}}V}+(\mbox{brane terms}) = 0. 
 \label{der_sgm_B}
\eea
We can see that constant $\sgm$ and $B$ is a trivial solution, 
and will focus on it in the following. 
Then, (\ref{EOM:V}) is rewritten as
\be
 \der_z\der_{\bar{z}}\ln\psi = -\frac{K}{2}\psi-2\pi\sum_k\alp_k\dlt^{(2)}(z-z_k), 
 \label{Liouville_eq}
\ee
where 
\be
 \psi \equiv \frac{\abs{\hc}^4}{b_{\rm H}^2\sgm^2}e^{-4cV}, \;\;\;\;\; 
 K \equiv \frac{2b_{\rm H}c^2}{f}, \;\;\;\;\;
 \alp_k \equiv \frac{c^2C_k\sqrt{b_{\rm H}}}{4\pi f\sqrt{\sgm}}. 
 \label{def:K}
\ee
This is the Liouville equation, and its solution can be expressed 
in the form of~\cite{Horvathy:1998pe,Redi:2004tm} 
\be
 \psi = \frac{4\abs{w'}^2}{K\brkt{1+\abs{w}^2}^2}, 
\ee
where $w(z)$ is a meromorphic function of $z$, and $w'\equiv dw/dz$. 
Noting that 
\be
 \der_z\der_{\bar{z}}\ln\abs{z}^2 = 2\pi\dlt^{(2)}(z), 
\ee
$\psi$ should behave near the brane locations as~\footnote{
When one of the branes is located at the infinity, we should use another coordinate patch, such as
$\tl{z}\equiv -1/z$, in order to describe it by the delta function. 
} 
\be
 \psi(z,\bar{z}) \sim \begin{cases} \abs{z-z_k}^{-2\alp_k} & (z \sim z_k) \\
 \abs{z}^{2\alp_\infty-4} & (\abs{z} \sim \infty) \end{cases}. 
 \label{asymp:psi}
\ee
We should note that there is an ambiguity in the expression of $w(z)$ for a given $\psi(z,\bar{z})$. 
In fact, $\psi$ does not change under the transformation,
\be
 w(z) \to \begin{pmatrix} a_{11} & a_{12} \\ a_{21} & a_{22} \end{pmatrix}\cdot w(z)
 \equiv \frac{a_{11}w(z)+a_{12}}{a_{21}w(z)+a_{22}},  \label{red_w}
\ee
where $a_{ij}$ ($i,j=1,2$) are complex constants, and 
\be
 \begin{pmatrix} a_{11} & a_{12} \\ a_{21} & a_{22} \end{pmatrix} \in {\rm SU(2)}.  
\ee
The asymptotic behavior~(\ref{asymp:psi}) is obtained when $w(z)$ behaves as 
\be
 w(z) \sim \begin{cases} \displaystyle M_k\cdot\brkt{z-z_k}^{1-\alp_k} & (z \sim z_k) \\
 M_\infty\cdot z^{1-\alp_\infty} & (\abs{z} \sim \infty)
 \end{cases} \label{asymp:w}
\ee
for $\alp_k<1$ and $M_k,M_\infty\in{\rm SU(2)}$. 

In summary, the background solution is
\bea
 &&\sgm = (\mbox{constant}), \;\;\;\;\;
 B = (\mbox{constant}), \;\;\;\;\;
 H_{\rm even} = \begin{pmatrix} \hc \\ 0 \\ \vdots \end{pmatrix}, \;\;\;\;\;
 H_{\rm odd} = \vec{0}, \nonumber\\
 &&V = -\frac{1}{4c}\ln\brc{\frac{2fb_{\rm H}\sgm^2}{c^2\abs{\hc}^4}\frac{\abs{w'}^2}{\brkt{1+\abs{w}^2}^2}}, 
 \;\;\;\;\;
 \Sgm = 0, \nonumber\\
 &&\cV_\rmT = \sgm\psi = \frac{\abs{\hc}^4}{b_{\rm H}^2\sgm}e^{-4cV}
 = \frac{4\sgm\abs{w'}^2}{K\brkt{1+\abs{w}^2}^2}. 
 \label{bgd:solution}
\eea

\subsection{Expressions in Wess-Zumino gauge}
Here we translate the background~(\ref{bgd:solution}) to the component-field expression 
in 6D SUGRA. 
As mentioned in the footnote~\ref{sf-comp:corr}, we need to move to 
the Wess-Zumino gauge for this purpose. 
This can be achieved by using the (super) gauge transformation for the background given by
\bea
 \tl{V} \eql V+\Lmd+\bar{\Lmd}, \;\;\;\;\;
 \tl{\Sgm} = \Sgm+\derE\Lmd = \Sgm+2\bar{\eta}\der_{\bar{z}}\Lmd, \nonumber\\
 \tl{H}_{\rm even} \eql e^{2c\Lmd}H_{\rm even}, \;\;\;\;\;
 \tl{H}_{\rm odd} = e^{-2c\Lmd}H_{\rm odd}, 
 \label{gauge_trf}
\eea
(and other superfields are neutral) with 
\be
 \Lmd = -\frac{V}{2} 
 = \frac{1}{8c}\ln\brc{\frac{2fb_{\rm H}\sgm^2}{c^2\abs{\hc}^4}\frac{\abs{w'}^2}{\brkt{1+\abs{w}^2}^2}}. 
 \label{choice_Lmd}
\ee
Then we have the background in this gauge as
\bea
 \tl{H}_{\rm even} \eql \brc{\frac{2fb_{\rm H}\sgm^2}{c^2\abs{\hc}^4}\frac{\abs{w'}^2}{\brkt{1+\abs{w}^2}^2}}^{1/4}
 \begin{pmatrix} \hc \\ 0 \\ \vdots \end{pmatrix}, \;\;\;\;\;
 \tl{H}_{\rm odd} = \vec{0}, \nonumber\\
 \tl{V} \eql 0, \;\;\;\;\;
 \tl{\Sgm} = \frac{\bar{\eta}}{4c}\brkt{\frac{\bar{w}''}{\bar{w}'}-\frac{2w\bar{w}'}{1+\abs{w}^2}}, 
 \nonumber\\
 \tl{\cV}_\rmT \eql \frac{4\sgm\abs{w'}^2}{K\brkt{1+\abs{w}^2}^2}, 
\eea
where $\sgm$ and $B$ are unchanged. 
Recalling that $\SE=\eta$ in our coordinates 
and comparing the above expressions with those in Appendix~\ref{comp_fields}, 
we obtain 
\bea
 \brkt{E_4E_5}^{1/4}\phi_2^2 
 \eql \brc{\frac{2fb_{\rm H}\sgm^2}{c^2\abs{\hc}^4}\frac{\abs{w'}^2}{\brkt{1+\abs{w}^2}^2}}^{1/4}\hc, \;\;\;\;\;
 \brkt{E_4E_5}^{1/4}\phi_2^{a\neq 2} =0, \nonumber\\
 \frac{i}{2}\brkt{A_4+iA_5} 
 \eql \frac{1}{4c}\brkt{\frac{\bar{w}''}{\bar{w}'}-\frac{2w\bar{w}'}{1+\abs{w}^2}}, \nonumber\\
 e^{(2)} \eql \psi = \frac{4\abs{w'}^2}{K\brkt{1+\abs{w}^2}^2}, \;\;\;\;\;
 B_{\udl{4}\udl{5}} = \frac{B}{4}. 
 \label{bgd:solution:15}
\eea
Notice that 
\be
 E_5 = iE_4, \;\;\;\;\;
 e^{(2)} = \Im\brkt{\bar{E}_4E_5} = \abs{E_4}^2,  
\ee
which follow from $\SE=\eta$, and 
\be
 E_4E_5(\phi_2^2)^4 = \frac{2fb_{\rm H}\sgm^2}{c^2}\frac{(\hc)^4}{\abs{\hc}^4}\frac{\abs{w'}^2}{\brkt{1+\abs{w}^2}^2}, 
\ee
where we have used (\ref{def:K}). 
Thus, the background can be expressed as
\bea
 ds^2 \eql \eta_{\mu\nu}dx^\mu dx^\nu
 +\abs{E_4 dz}^2, \nonumber\\
 E_4 \eql -iE_5 = \frac{\sqrt{2f}\abs{w'}}{\sqrt{b_{\rm H}}c\brkt{1+\abs{w}^2}}
 \exp\brc{i\brkt{2\arg(\hc)-\frac{\pi}{4}}},  \nonumber\\
 A_4 \eql \frac{1}{2c}\Im\brkt{\frac{\bar{w}''}{\bar{w}'}-\frac{2w\bar{w}'}{1+\abs{w}^2}}, \;\;\;\;\;
 A_5 = -\frac{1}{2c}\Re\brkt{\frac{\bar{w}''}{\bar{w}'}-\frac{2w\bar{w}'}{1+\abs{w}^2}}, \nonumber\\
 \phi_2^2 \eql \brkt{b_{\rm H}\sgm}^{1/2}, \;\;\;\;\;
 \phi_2^{a\neq 2} = 0, \nonumber\\
 \sgm \eql (\mbox{real constant}), \;\;\;\;\;
 B_{45} = e^{(2)}B_{\udl{4}\udl{5}} = \frac{B\abs{w'}^2}{K\brkt{1+\abs{w}^2}^2}, 
 \label{bgd:solution:2}
\eea
where the constants~$f$, $b_{\rm H}$, $c$ and $B$ are real, 
and $\hc$ is complex.  

Since the background metric for the compact space is 
\be
 \abs{E_4}^2dzd\bar{z} = e^{(2)}dzd\bar{z}
 = \psi dzd\bar{z}, 
\ee
and $\psi$ behaves as (\ref{asymp:psi}) near the singularities, 
the space has the conical singularities at $z=z_k$, and 
$\alp_k$ defined in (\ref{def:K}) can be identified with the deficit angles, 
which are proportional to the brane tensions~\cite{Redi:2004tm}.\footnote{
In the Planck unit, the tension~$\tau_k$ is equal to $2\pi\alp_k$. 
} 
Besides, the volume of the compact space is given by 
\be
 {\rm Vol}^{(2)} = \int d x^4dx^5\;e^{(2)} = \frac{1}{2}\int d^2z\;e^{(2)} 
 = \frac{1}{2}\int d^2z\;\psi. 
\ee
In order for this integral to have a finite value, 
(\ref{asymp:psi}) indicates that $\alp_k<1$ must be satisfied for all $k$. 
Using the Gauss-Bonnet formula, this integral is calculated as~\cite{Redi:2004tm}
\be
 {\rm Vol}^{(2)} = \frac{2}{K}\int\dr^2z\;\frac{\abs{w'}^2}{\brkt{1+\abs{w}^2}^2} 
 = \frac{2\pi}{K}\brkt{2-2g-\sum_k\alp_k}, 
 \label{volume}
\ee
where $g$ is the genus of the compact space.

\subsection{Localized fluxes and total flux}
After moving to the Wess-Zumino gauge, there still remains the gauge degree of freedom. 
We can add an arbitrary imaginary part of $\Lmd$ to (\ref{choice_Lmd}) 
maintaining the background~$\tl{V}=0$. 
In such gauges, (the extra-dimensional components of) the gauge potential is expressed 
as $A_{\bar{z}}=-i\eta\tl{\Sgm}=-2i\der_{\bar{z}}\Lmd$. 
Thus, the field strength~$F_{z\bar{z}}$ is 
\be
 F_{z\bar{z}} \equiv \der_z A_{\bar{z}}-\der_{\bar{z}}A_z 
 = -2i\der_z\der_{\bar{z}}\brkt{\Lmd+\bar{\Lmd}}, 
\ee
which is certainly gauge-invariant under the remaining gauge transformation.\footnote{
The fact that this is not super-gauge invariant reflects 
the fact that we cannot construct a field-strength superfield that contains $F_{z\bar{z}}$. 
}
Using (\ref{choice_Lmd}) and (\ref{Liouville_eq}), $F_{z\bar{z}}$ is calculated as
\bea
 F_{z\bar{z}} \eql 2i\der_z\der_{\bar{z}}V
 = \frac{iK}{4c}\psi+\frac{i\pi}{c}\sum_k\alp_k\dlt^{(2)}(z-z_k) \nonumber\\
 \eql \frac{i\abs{w'}^2}{c\brkt{1+\abs{w}^2}^2}
 +\frac{i\pi}{c}\sum_k\alp_k\dlt^{(2)}(z-z_k). 
 \label{expr:Fzbz}
\eea
Therefore, the brane terms~(\ref{L_brane}) induce the brane-localized fluxes. 
The total flux is calculated using (\ref{volume}) as~\footnote{
If we choose $H_{\rm odd}^1=H^1$ as the brane compensator in (\ref{L_brane}) 
and as the only non-vanishing background 
among $H_{\rm odd}$ and $H_{\rm even}$, the total flux becomes $\cB=-\pi(2-2g)$. 
}   
\bea
 \cB \eql \int dx^4dx^5\;cF_{45} = -i\int d^2z\;cF_{z\bar{z}} 
 = \int d^2z\;\frac{\abs{w'}^2}{\brkt{1+\abs{w}^2}^2}+\pi\sum_k\alp_k \nonumber\\
 \eql \pi\brkt{2-2g-\sum_k\alp_k}+\pi\sum_k\alp_k
 = \pi\brkt{2-2g}. 
 \label{total_flux}
\eea
We have used that $\int dx^4dx^5=\frac{1}{2}\int d^2z$, $F_{z\bar{z}}=-2iF_{45}$, and (\ref{volume}). 
Thus, the total flux~$\cB$ is independent of the brane tensions. 
Eq.(\ref{total_flux}) indicates that 
the background solution~(\ref{bgd:solution:2}) automatically satisfies the flux quantization condition 
in Appendix~\ref{FQ:sphere}. 

So far, we have not specified the compact space. 
The form of $w(z)$ in (\ref{bgd:solution}) or (\ref{bgd:solution:2}) depends on it. 
In the component-field expressions, this issue is discussed 
in the previous works~\cite{Redi:2004tm,Akerblom:2009ev,Akerblom:2010xb}. 
In the next two sections, we consider specific compactifications 
and summarize those results in our notations, 
for the sake of completeness.

\section{Sphere compactification} \label{sphere_cmp}
Let us consider the case that the superfields are defined in the entire complex plane including infinity, 
\ie, the Riemann sphere. 
In this case, the tensions are constrained as
\be
 \alp_k <1, \;\;\;\;\;
 \sum_k \alp_k < 2. 
\ee
The second one comes from the condition that the volume~(\ref{volume}) should be positive. 

\subsection{In the absence of branes}
In the absence of the branes, $w(z)$ has no singularities and is holomorphic over the whole complex plane. 
Thus we can redefine the complex coordinate as $z\to\tl{z}\equiv w(z)$, 
and obtain 
\be
 ds_2^2 = \frac{4d\tl{z}d\bar{\tl{z}}}{K\brkt{1+\abs{\tl{z}}^2}^2}. 
\ee
This is nothing but the Fubini-Study metric. 
Hence the compactified space is a sphere with the radius~$1/\sqrt{K}$. 
In this case, the background~(\ref{bgd:solution:2}) represents 
the Salam-Sezgin solution~\cite{Salam:1984cj}.\footnote{
The constant~$B$ is chosen to zero in Ref.~\cite{Salam:1984cj}. 
}

\subsection{In the presence of branes}
In the presence of the branes, the solution of (\ref{Liouville_eq}) is found 
by using the technology of the fuchsian equations~\cite{Redi:2004tm}. 
In this case, $w(z)$ is given by 
\be
 w(z) = \frac{u_1(z)}{u_2(z)}, 
\ee
where $u_1(z)$ and $u_2(z)$ are two linearly independent solutions of the fuchsian equation, 
\be
 \frac{d^2u}{dz^2}+\sum_{k=1}^{N-1}\brc{\frac{\alp_k(2-\alp_k)}{4(z-z_k)^2}+\frac{\bt_k}{2(z-z_k)}}u = 0. 
 \label{fuchs_eq}
\ee
The constants~$\bt_k$ are known as the accessory parameters. 
The condition that $\abs{z}=\infty$ is a regular singular point requires
\bea
 &&\sum_{k=1}^{N-1}\bt_k = 0, \;\;\;\;\;
 \sum_{k=1}^{N-1}\brc{2\bt_kz_k+\alp_k(2-\alp_k)} = \alp_\infty (2-\alp_\infty), \nonumber\\
 &&\sum_{k=1}^{N-1}\brc{\bt_k z_k^2+z_k\alp_k(2-\alp_k)} = \bt_\infty. 
\eea
Thus only $N-3$ parameters among $\bt_k$ are independent. 
%Near the singular points~$z\sim z_k$, a solution of (\ref{fuchs_eq}) behaves as
%\be
% u(z) \sim A_{k1}(z-z_k)^{1-\frac{\alp_k}{2}}+A_{k2}(z-z_k)^{\frac{\alp_k}{2}}, 
%\ee
%where $A_{k1}$ and $A_{k2}$ are complex constants. 
Going around the singularity~$z=z_k$, the two solutions transform as 
\be
 \begin{pmatrix} u_1(z) \\ u_2(z) \end{pmatrix} 
 \to M_k\begin{pmatrix} u_1(z) \\ u_2(z) \end{pmatrix}, 
\ee
where the monodromy matrix~$M_k$ generically belongs to SL(2,$\mathbb C$). 
Then, $w(z)$ transforms as
\be
 w(z) \to M_k\cdot w(z), 
\ee
where the operation of the matrix~$M_k$ is defined in (\ref{red_w}). 
Hence, in order for $\psi(z,\bar{z})$ to be single-valued on the complex plane, 
we have to choose the two independent solutions~$u_1(z)$ and $u_2(z)$ such that 
$M_k\in {\rm SU(2)}$. 

As the simplest example, consider the case that the origin~$z=0$ is the only singularity on the complex plane. 
In this case, (\ref{fuchs_eq}) becomes 
\be
 \frac{d^2u}{dz^2}+\frac{\alp_1(2-\alp_1)}{4z^2}u = 0. 
\ee 
If we choose the two independent solutions as
\be
 u_1(z) = z^{1-\frac{\alp_1}{2}}, \;\;\;\;\;
 u_2(z) = z^{\frac{\alp_1}{2}}, 
\ee 
the monodromy matrix becomes $M_1=\diag(e^{-\pi i\alp_1},e^{\pi i\alp_1})$, 
which belongs to SU(2). 
Thus, the desired background is obtained by
\bea
 w(z) \eql \frac{u_1(z)}{u_2(z)} = z^{1-\alp_1}, \nonumber\\
 \psi(z,\bar{z}) \eql \frac{4\abs{w'}^2}{K\brkt{1+\abs{w}^2}^2} 
 = \frac{4(1-\alp_1)^2\abs{z}^{-2\alp_1}}{K\brkt{1+\abs{z}^{2-2\alp_1}}^2}. 
 \label{rugby}
\eea
Recall that $\alp_1<1$ from the requirement that (\ref{volume}) is finite. 
So the asymptotic behavior of $\psi(z,\bar{z})$ for $\abs{z}\gg 1$ is
\be
 \psi(z,\bar{z}) \sim \frac{4(1-\alp_1)^2}{K}\abs{z}^{2\alp_1-4}, 
\ee
which indicates that the infinity is also a singular point with $\alp_\infty=\alp_1$ from (\ref{asymp:psi}). 
Therefore, there are at least two singularities in the case of the sphere compactification 
in the presence of the branes. 
This is in contrast to the torus compactification (see Sect.~\ref{Olesen_sol}).  
The background solution with (\ref{rugby}) represents 
the so-called rugby-ball (or football) solution~\cite{Carroll:2003db,Navarro:2003vw,Navarro:2003bf}. 
For the case with more branes, see Ref.~\cite{Redi:2004tm}. 

The author of Ref.~\cite{Redi:2004tm} focuses on the case that $0<\alp_k<1$. 
However, this condition can be released as $\alp_k<1$ once negative-tension branes are accepted, 
just like in the Randall-Sundrum model~\cite{Randall:1999ee}. 
Especially, when all $\alp_k$ are integers, 
the rational functions are allowed as $w(z)$~\cite{Horvathy:1998pe}.

\section{Torus compactification} \label{torus_cmp}
Now we consider the case that the extra dimensions are compactified on a torus. 
The points are identified as
\be
 z \sim \tl{z}_{m,n} \equiv z+m+n\tau, \;\;\;\;\; (m,n\in {\mathbb Z})
\ee
where $\tau$ is a complex constant, and $\Im\tau>0$. 
Since $\psi$ is proportional to $e^{(2)}$, it satisfies the periodic boundary conditions, 
\be
 \psi(\tl{z}_{m,n},\bar{\tl{z}}_{m,n}) = \psi(z,\bar{z}). \label{BC:psi:torus}
\ee
Recalling the redundancy of $w(z)$ under (\ref{red_w}), they are satisfied 
when $w(z)$ is subject to the boundary conditions, 
\bea
 w(z+1) \eql \gm_1\cdot w(z), \nonumber\\
 w(z+\tau) \eql \gm_\tau\cdot w(z),  \label{tsl:torus}
\eea
where
\be
 \gm_1, \, \gm_\tau \in {\rm SU(2)}. 
\ee
The matrices~$\gm_1$ and $\gm_\tau$ either commute or anticommute 
to each other~\cite{Akerblom:2009ev}.

\subsection{In the absence of branes}
In the absence of the brane terms, \ie, $C_k=0$, 
there is no solution to the Liouville equation~(\ref{Liouville_eq}) 
that satisfies the boundary conditions in (\ref{BC:psi:torus}), 
unless $c=0$. 
When $c=0$, \ie, 6D SUGRA is not gauged, 
we have a constant solution, 
\be
 \psi = (\mbox{constant}). 
\ee
Namely, $V$ and $\cV_\rmT$ in (\ref{bgd:solution}) should be replaced with 
\be
 V = (\mbox{constant}), \;\;\;\;\;
 \cV_\rmT = \frac{\abs{\hc}^4}{b_{\rm H}^2\sgm}. 
\ee

\subsection{Olesen solution} \label{Olesen_sol}
Next we consider the case with branes. 
The constraints on the tensions in this case become
\be
 \alp_k < 1, \;\;\;\;\;
 \sum_k\alp_k < 0. 
\ee
Therefore, negative-tension branes are necessary. 
The solutions in this case are expressed by using the Weierstrass elliptic functions 
whose definitions are collected in Appendix~\ref{def:Welliptic}. 

In contrast to the sphere compactification, 
there is a solution with one brane, which is found in Ref.~\cite{Akerblom:2010xb}. 
Since the expression of the solution becomes simple for a square torus, 
we consider the case of $\tau=i$ in this subsection. 
The solution is given by
\be
 w(z) = \frac{\wp(z)}{\wp(1)}, \label{Olesen:wp}
\ee
where $\wp(z)\equiv \wp_{2,2i}(z)$ is the Weierstrass p-function. 
This leads to 
\be
 e^{(2)} = \psi 
 = \frac{4\abs{\wp(1)}^2\abs{\wp'(z)}^2}{K\brkt{\abs{\wp(1)}^2+\abs{\wp(z)}^2}^2}. 
 \label{Olesen:psi}
\ee
The corresponding compact space is called the Olesen space~\cite{Olesen:1991df,Olesen:1991dg}. 

From the definition of $\wp(z)$ in (\ref{def:wp}), we obtain the relation, 
\be
 \wp(i) = -\wp(1). 
\ee
Thus, we find from (\ref{formulae:wp}) that 
\bea
 w(z+1) \eql 1+\frac{2}{w(z)-1} = \frac{w(z)+1}{w(z)-1}, \nonumber\\
 w(z+i) \eql -1+\frac{2}{w(z)+1} = \frac{-w(z)+1}{w(z)+1}. 
\eea
Namely, the matrices~$\gm_1$ and $\gm_\tau$ in (\ref{tsl:torus}) are read off as
\be
 \gm_1 = \frac{i}{\sqrt{2}}\begin{pmatrix} 1 & 1 \\ 1 & -1 \end{pmatrix}, \;\;\;\;\;
 \gm_\tau = \frac{i}{\sqrt{2}}\begin{pmatrix} -1 & 1 \\ 1 & 1 \end{pmatrix}. 
 \label{gms:Olesen}
\ee
Since these belong to SU(2), (\ref{Olesen:psi}) is doubly periodic. 
\be
 \psi(z+1) = \psi(z+i) = \psi(z). 
\ee
As mentioned in Ref.~\cite{Akerblom:2010xb}, this compact space has 
the positive constant curvature~$K$ almost everywhere except for the origin~$z=0$. 
From the definition of $\wp(z)$ in (\ref{def:wp}), we can see that
\be
 w(z) \sim \frac{1}{z^2}, \;\;\;\;\;
 \psi(z,\bar{z}) \sim \abs{z}^2, 
\ee
near the origin. 
This indicates from (\ref{asymp:psi}) that there is 
a conical singularity at $z=0$ with the negative deficit angle (or tension)~$\alp_1=-1$.

\subsection{General solution with branes}
The solutions with an arbitrary number of the branes 
can be expressed by the two special solutions~$f_{\vph_1,\vph_\tau}(z)$ 
and $g(z)$, which satisfy 
\bea
 f_{\vph_1,\vph_\tau}(z+1) \eql e^{i\vph_1}f_{\vph_1,\vph_\tau}(z), \;\;\;\;\;
 f_{\vph_1,\vph_\tau}(z+\tau) = e^{i\vph_\tau}f_{\vph_1,\vph_\tau}(z), \nonumber\\
 g(z+1) \eql -g(z), \;\;\;\;\;
 g(z+\tau) = \frac{1}{g(z)}, 
\eea
where $\vph_1$ and $\vph_\tau$ are real constants~\cite{Akerblom:2009ev}. 
Here, $f_{\vph_1,\vph_\tau}(z)$ is called the elliptic function of the second kind 
with multipliers of unit modulus, and explicitly given by 
\be
 f_{\vph_1,\vph_\tau}(z) = \brc{s_0+\sum_{k=1}^Ns_k\frac{d^k\zeta}{dz^k}(z-z_0)}
 \frac{\sgm^N(z-z_0)}{\prod_{k=1}^N\sgm(z-z_k)}e^{\lmd z},  \label{def:f}
\ee
where $\zeta(z)\equiv\zeta_{1,\tau}(z)$ and $\sgm(z)\equiv\sgm_{1,\tau}(z)$ 
are the Weierstrass zeta- and sigma-functions, and 
\bea
 \lmd \defa \frac{\vph_\tau}{\pi}\zeta\brkt{\frac{1}{2}}
 -\frac{\vph_1}{\pi}\zeta\brkt{\frac{\tau}{2}}, \nonumber\\
 z_0 \defa \frac{1}{2\pi N}\brkt{\vph_\tau-\vph_1\tau}+\frac{1}{N}\sum_{k=1}^Nz_k. 
\eea
The constants~$s_0,s_1,\cdots,s_N$ and $z_1,\cdots,z_N$ are
\bea
 s_0,s_1,\cdots,s_N &\in& {\mathbb C}, \nonumber\\
 z_1,\cdots,z_N &\in& \brc{t_1+t_2\tau|0\leq t_1,t_2\leq 1}. 
\eea
The other special solution~$g(z)$ is given by
\be
 g(z) \equiv -\frac{f_{0,\pi}(z)-1}{f_{0,\pi}(z)+1}\cdot\frac{\wp(z)+b_0}{c_0\wp(z)+d_0}, 
\ee
where $\wp(z)\equiv\wp_{2,2\tau}(z)$, and 
\bea
 b_0 \defa \frac{-e_2+c_0^2(-2e_1+e_2)}{1+c_0^2}, \;\;\;\;\;
 d_0 \equiv \frac{c_0\brkt{-2e_1+e_2-c_0^2e_2}}{1+c_0^2}, \nonumber\\
 c_0 \defa \sqrt{\frac{-3e_1+2\sqrt{(e_1-e_2)(2e_1+e_2)}}{e_1+2e_2}}, 
\eea
with $e_1\equiv\wp(1)$, $e_2\equiv\wp(\tau)$, and $e_3\equiv -e_1-e_2$. 

The matrices~$\gm_1$ and $\gm_\tau$ for $f_{\vph_1,\vph_\tau}(z)$ are 
\be
 \gm_1 = \begin{pmatrix} e^{i\vph_1/2} & 0 \\ 0 & e^{-i\vph_1/2} \end{pmatrix}, \;\;\;\;\;
 \gm_\tau = \begin{pmatrix} e^{i\vph_\tau/2} & 0 \\ 0 & e^{-i\vph_\tau/2} \end{pmatrix}, 
\ee
which commute, 
while those for $g(z)$ are 
\be
 \gm_1 = \begin{pmatrix} -i & 0 \\ 0 & i \end{pmatrix}, \;\;\;\;\;
 \gm_\tau = \begin{pmatrix} 0 & i \\ i & 0 \end{pmatrix}, 
\ee
which anti-commute. 

In terms of the above special solutions, $w(z)$ is expressed as
\be
 w(z) = \begin{cases} U\cdot f_{\vph_1,\vph_\tau}(z) & (\mbox{$\gm_1$ and $\gm_\tau$ commute}) \\
 U\cdot g(z) & (\mbox{$\gm_1$ and $\gm_\tau$ anti-commute}) \end{cases}, 
 \label{expr:w:2}
\ee
where $U\in{\rm U(2)}$ is a constant matrix. 
Since the matrices in (\ref{gms:Olesen}) anti-commute to each other, 
the Olesen solution~(\ref{Olesen:wp}) can be expressed in the form of $U\cdot g(z)$.

\section{Summary} \label{summary}
We provided $\cN=1$ superfield description of BPS backgrounds that preserve $\cN=1$ SUSY 
in 6D SUGRA compactified on a sphere or torus, including brane-localized FI terms. 
It is obtained by solving the background superfield EOMs, which are derived 
from the superfield action in our previous paper~\cite{Abe:2017pvw}. 
%Since the relation between the superfields and the usual component-field expressions is 
%somewhat complicated as shown in Appendix~\ref{comp_fields}, 
We found that the gauge in which $\vev{\SE}=e^{-\pi i/4}$ and $\vev{\Sgm}=0$ 
is convenient to solve the background EOMs. 
We focused on the case that the dilaton~$\sgm$ and the tensor component~$B_{\udl{4}\udl{5}}$ 
have constant background, which corresponds to the unwarped geometry of the extra dimensions. 
One of the background equations in this case is reduced to the Liouville equation, 
whose solutions have been well-investigated. 
Moving to the Wess-Zumino gauge for the gauge superfield~$V$, 
we can read off the component-field expression of the background 
by using the expressions listed in Appendix~\ref{comp_fields}. 
The background obtained in this paper reproduces the known results 
in the previous works. 
The brane terms in (\ref{L_brane}) induce the localized magnetic flux in the extra dimensions, 
keeping the total flux unchanged. 

We can also consider the warped geometry by looking for non-constant solutions of (\ref{der_sgm_B}). 
Such a case was discussed in the component-field expressions in Ref.~\cite{Lee:2005az}. 
Because of their complicated definition of the coordinates, it is a nontrivial task to express their solution 
in our superfield language. 
Our results help us develop a systematic derivation of 4D effective action from 
the superfield action of 6D SUGRA with the brane terms. 
These issues will be discussed in our subsequent papers.

\subsection*{Acknowledgements}
This work was supported in part by JSPS KAKENHI Grant Numbers JP16K05330 (H.A.), 
JP16J06569 (S.A.) and JP25400283 (Y.S.).

\appendix

\section{Correspondence to the component fields} \label{comp_fields}
Here we provide some of the components of the $\cN=1$ superfields in Sect.~\ref{superfields} 
in terms of the component fields in 6D SUGRA~\cite{Abe:2015yya,Abe:2017pvw,Sakamura:2011df}. 

\subsection{Weyl multiplet}
The 6D Weyl multiplet~$\bE$ consists of the sechsbein~$e_M^{\;\;\udl{N}}$, 
the gravitinos~$\psi^i_{M\bar{\alp}}$, and the auxiliary  fields, 
where $\bar{\alp}$ is a 6D spinor index, and $i,j=1,2$ are the $\suU$-doublet indices. 
The gravitino has the 6D chirality~$+$, and is the $\suU$-Majorana-Weyl fermion, 
which can be decomposed into the two 4D Dirac fermions as 
\be
 \psi_M^1 = \begin{pmatrix} \psi_{M\alp}^+ \\ \bar{\psi}_M^{-\dalp} \end{pmatrix}, \;\;\;\;\;
 \psi_M^2 = \begin{pmatrix} -\psi_{M\alp}^- \\ \bar{\psi}_M^{+\dalp} \end{pmatrix}, 
\ee
where $\alp,\dalp=1,2$ are the 2-component spinor indices. 
Then the components of the superfields in (\ref{sf:Weyl}) are expressed as
\bea
 U^\mu \eql (\tht\sgm^{\udl{\nu}}\bar{\tht})\tl{e}_\nu^{\;\;\mu}
 +i\bar{\tht}^2\brkt{\tht\sgm^{\udl{\nu}}\bar{\sgm}^{\udl{\mu}}\psi_{\udl{\nu}}^+}
 -i\tht^2\brkt{\bar{\tht}\bar{\sgm}^{\udl{\nu}}\sgm^{\udl{\mu}}\bar{\psi}_{\udl{\nu}}^+}+\cdots, \nonumber\\
 U^m \eql \brkt{\tht\sgm^{\udl{\mu}}\bar{\tht}}e_{\udl{\mu}}^{\;\;m}
 -\bar{\tht}^2\brkt{\tht\sgm^{\udl{\mu}}\bar{\psi}_{\udl{\mu}}^-}\brkt{e_{\udl{4}}^{\;\;m}+ie_{\udl{5}}^{\;\;m}}
 +\tht^2\brkt{\bar{\tht}\bar{\sgm}^{\udl{\mu}}\psi_{\udl{\mu}}^-}
 \brkt{e_{\udl{4}}^{\;\;m}-ie_{\udl{5}}^{\;\;m}} \nonumber\\
 &&+\cdots, \nonumber\\
 \Psi_{m\alp} \eql \frac{i}{2}\brkt{\sgm^{\udl{\mu}}\bar{\tht}}_\alp e_{m\udl{\mu}}+\cdots, \nonumber\\
 \SE \eql \sqrt{\frac{E_4}{E_5}}+\cdots, \nonumber\\
 \VE \eql e^{(2)}+\cdots, 
 \label{comp:U^mu}
\eea
where the ellipses denote higher order terms in $\tht$ or $\bar{\tht}$, and
\bea
 \tl{e}_\nu^{\;\;\mu} \defa e_\nu^{\;\;\udl{\mu}}-\vev{e_\nu^{\;\;\udl{\mu}}}
 = e_\nu^{\;\;\udl{\mu}}-\dlt_\nu^{\;\;\mu}, \nonumber\\
 E_m \defa e_m^{\;\;\udl{4}}+ie_m^{\;\;\udl{5}}, \nonumber\\
 e^{(2)} \defa \det (e_m^{\;\;\udl{n}}) = e_4^{\;\;\udl{4}}e_5^{\;\;\udl{5}}-e_4^{\;\;\udl{5}}e_5^{\;\;\udl{4}}. 
 \label{def:E_m}
\eea
Note that we need not discriminate the 4D flat and curved indices for $\tl{e}_\nu^{\;\;\mu}$ 
at the linearized order since $\vev{e_\nu^{\;\;\udl{\mu}}}=\dlt_\nu^{\;\;\mu}$. 

Thus, the lowest component of $R_{\rm E}^-$ is calculated as
\be
 R_{\rm E}^- = \frac{e^{(2)}}{\abs{E_4E_5}}.  \label{lowest:R_E}
\ee

\subsection{Hypermultiplet}
The hypermultiplet~$\bH^A$ consists of the hyperscalar~$(\phi_i^{2A-1},\phi_i^{2A})$, 
which are subject to the reality condition:~$(\phi_1^{2A-1})^*=\phi_2^{2A}$, 
$(\phi_1^{2A})^*=-\phi_2^{2A-1}$, the hyperino~$(\zeta^{2A-1}_{\bar{\alp}},\zeta^{2A}_{\bar{\alp}})$, 
which are the symplectic Majorana spinors, 
and the auxiliary fields. 
The lowest components of $H^{\bar{A}}$ in (\ref{sf:hyper}) are given by 
\be
 H^{2A-1} = (E_4E_5)^{1/4}\phi^{2A-1}_2+\cdots, \;\;\;\;\;
 H^{2A} = (E_4E_5)^{1/4}\phi^{2A}_2+\cdots, 
 \label{comp:hyper}
\ee
where $E_m$ ($m=4,5$) are defined in (\ref{def:E_m}).

\subsection{Vector multiplet}
The vector multiplet~$\bV^I$ consists of the gauge field~$A_M^I$, 
the gauginos~$\Omg_{\bar{\alp}}^{Ii}$ and the auxiliary fields. 
The gauge field is embedded into the superfields as
\bea
 V^I \eql -\brkt{\tht\sgm^{\udl{\mu}}\bar{\tht}}A_{\udl{\mu}}^I+\cdots, \nonumber\\
 \Sgm^I \eql \frac{i}{2}\brkt{\frac{1}{\SE|}A_4^I-\SE|A_5^I}+\cdots, 
 \label{comp:V^I}
\eea
where $\SE|=\sqrt{E_4/E_5}$ is the lowest component of $\SE$.

\subsection{Tensor multiplet}
The tensor multiplet~$\bT$ consists of a real scalar~$\sgm$, 
an anti-symmetric tensor field~$B_{MN}$, the fermionic fields and the auxiliary fields. 
They are embedded into the superfields as
\bea
 \Ups_{\rmT\alp} \eql -\tht_\alp\brkt{2B_{\udl{4}\udl{5}}+i\sgm}
 -2i\brkt{\sgm^{\udl{\mu}\udl{\nu}}\tht}_\alp B_{\udl{\mu}\udl{\nu}}+\cdots, \nonumber\\
 V_{\rmT m} \eql -2\brkt{\tht\sgm^{\udl{\mu}}\bar{\tht}}B_{\udl{\mu}m}+\cdots, \nonumber\\
 \Sgm_{\rmT} \eql \frac{e^{(2)}}{2}\sgm-iB_{45}+\cdots. 
 \label{comp:Ups_T}
\eea

From these expressions, 
we can calculate the lowest components of the field-strength superfields in $\bT$ as
\be
 \cX_{\rmT} = \sgm+\cdots, \;\;\;\;\;
 \cV_\rmT = e^{(2)}\sgm+\cdots. \label{comp:cV_T}
\ee

\section{Background equations of motion for $\bdm{U^m}$} \label{bgd_eq:U^m}
Picking up the linear terms in $U^m$ from the Lagrangian terms in Sect.~\ref{inv_action}, we have 
\bea
 &&-2\abs{J_\cP}\brkt{\frac{\cV_\rmT R_{\rm E}^-}{\cX_\rmT}}^{1/2}
 \brkt{\hat{H}_{\rm odd}^\dagger\tl{d}e^{V}\hat{H}_{\rm odd}
 +\hat{H}_{\rm even}^\dagger\tl{d}e^{-V}\hat{H}_{\rm even}} \nonumber\\
 \toa 2\brkt{\frac{R_{\rm E}^-}{\cV_\rmT\cX_\rmT}}^{1/2}L_{\rm H}
 \brkt{U^m\Im\der_m\Sgm_\rmT+\der_m U^m\Im\Sgm_\rmT} \nonumber\\
 &&+\brkt{\frac{\cV_\rmT}{\cX_\rmT R_{\rm E}^-}}^{1/2}L_{\rm H}
 \brc{2U^m\Re\frac{\bar{S}_{\rm E}\der_m\SE}{\SE^2}
 +\brkt{\der_4 U^4-\der_5 U^5}R_{\rm E}^+
 +\frac{\der_4U^5}{\abs{\SE}^2}-\abs{\SE}^2\der_5 U^4} \nonumber\\
 &&-\frac{1}{2}U^m\brkt{\frac{\cV_\rmT R_{\rm E}^-}{\cX_\rmT^3}}^{1/2}L_{\rm H}
 \Re\brkt{\der_m D^\alp\Ups_{\rmT\alp}} \nonumber\\
 &&+4U^m\brkt{\frac{\cV_\rmT R_{\rm E}^-}{\cX_\rmT}}^{1/2}
 \Im\brkt{H_{\rm odd}^\dagger\tl{d}e^{V}\der_m H_{\rm odd}
 +H_{\rm even}^\dagger\tl{d}e^{-V}\der_m H_{\rm even}}, 
\eea
where $L_{\rm H}$ is defined in (\ref{def:L_H}), 
\bea
 &&J_\cP f_{IJ}\brc{-2\hat{\Sgm}^I D^{\cP\alp}V^I\hat{\cY}_{\rmT\alp}
 +\frac{1}{2}\brkt{\derE^\cP V^ID^{\cP\alp}V^J-\derE^\cP D^{\cP\alp}V^IV^J}\hat{\cY}_{\rmT\alp}} 
 \nonumber\\
 \toa f_{IJ}\bigg\{2iU^m\der_m\brkt{\Sgm^IV^J} \nonumber\\
 &&\hspace{7mm}\left.
 +\frac{1}{2}\brkt{-iU^m\derE V^I\der_m V^J+iU^m\derE\der_m V^IV^J
 +i\derE U^m\der_m V^IV^J}\right\}D^\alp\cY_{\rmT\alp} \nonumber\\
 &&+2iU^m f_{IJ}\Sgm^IV^J\der_m D^\alp\cY_{\rmT\alp}+\cdots, 
\eea
and 
\bea
 &&\frac{\cX_{\rmT}}{R_{\rm E}^-}f_{IJ}\left\{
 4\brkt{\derE^\cP V^I-\hat{\Sgm}^I}^\dagger\brkt{\derE^\cP V^J-\hat{\Sgm}^J}
 -2\brkt{\derE^\cP V^I}^\dagger\derE^\cP V^J
 +\brkt{2J_\cP R_{\rm E}^+\hat{\Sgm}^I\hat{\Sgm}^J+\Hc}\right\} \nonumber\\
 \toa \frac{L_{\rm V}}{2R_{\rm E}^-}U^m\Re\brkt{\der_m D^\alp\Ups_{\rmT\alp}} 
 -\frac{\cX_\rmT L_{\rm V}}{R_{\rm E}^{-2}}\dlt R_{\rm E}^- \nonumber\\
 &&+\frac{\cX_\rmT f_{IJ}}{R_{\rm E}^-}\bigg[
 2i\brkt{\bar{\der}_{\rm E}U^m\der_m V^I+U^m\der_m\bar{S}_{\rm E}\bar{\cO}_{\rm E}V^I}
 \brkt{\derE V^J-2\Sgm^J}+4iU^m\der_m\bar{\Sgm}^I\brkt{\derE V^J-\Sgm^J} \nonumber\\
 &&\hspace{17mm}
 +2iR_{\rm E}^+\der_m\brkt{U^m\Sgm^I\Sgm^J}
 +2(\dlt R_{\rm E}^+)\Sgm^I\Sgm^J+\Hc\bigg], 
\eea
where
\bea
 \dlt R_{\rm E}^- \defa -2U^m\Re\frac{\bar{S}_{\rm E}\der_m\SE}{\SE^2}
 -\brkt{\der_4 U^4-\der_5 U^5}R_{\rm E}^+-\frac{\der_4 U^5}{\abs{\SE}^2}
 +\abs{\SE}^2\der_5 U^4, \nonumber\\
 \dlt R_{\rm E}^+ \defa 2U^m\Im\frac{\bar{S}_{\rm E}\der_m\SE}{\SE^2}
 +\brkt{\der_4 U^4-\der_5 U^5}R_{\rm E}^+-\frac{i\der_4 U^5}{\abs{\SE}^2}-i\abs{\SE}^2\der_5U^4, 
 \nonumber\\
 L_{\rm V} \defa f_{IJ}\left\{4\brkt{\derE V^I-\Sgm^I}^\dagger\brkt{\derE V^J-\Sgm^J}
 -2\brkt{\derE V^I}^\dagger\derE V^J \right. \nonumber\\
 &&\hspace{10mm}\left.
 +2\Re\frac{\bar{S}_{\rm E}}{\SE}\brkt{\Sgm^I\Sgm^J+\Hc}\right\}. 
 \label{def:L_V}
\eea
The ellipsis denotes terms involving the spinor derivative of the superfields other than $\Ups_{\rmT\alp}$ 
and $\cY_{\rmT\alp}$. 
The other Lagrangian terms give no contributions to the background EOMs. 
 
Then we can read off the equation of motion for $U^4$ as
\bea
 0 \eql -2\Im\Sgm_\rmT\der_4\brc{\brkt{\frac{R_{\rm E}^-}{\cV_\rmT\cX_\rmT}}^{1/2}L_{\rm H}}
 +2\brkt{\frac{\cV_\rmT}{\cX_\rmT R_{\rm E}^-}}^{1/2}L_{\rm H}\Re\frac{\bar{S}_{\rm E}\der_4\SE}{\SE^2} 
 \nonumber\\
 &&-\frac{1}{2}\brkt{\frac{\cV_\rmT R_{\rm E}^-}{\cX_{\rm T}^3}}^{1/2}L_{\rm H}
 \Re\brkt{\der_4 D^\alp\Ups_{\rmT\alp}} \nonumber\\
 &&+4\brkt{\frac{\cV_\rmT R_{\rm E}^-}{\cX_\rmT}}^{1/2}
 \Im\brkt{H_{\rm odd}^\dagger\tl{d}e^{V}\der_4 H_{\rm odd}
 +H_{\rm even}^\dagger\tl{d}e^{-V}\der_4 H_{\rm even}} \nonumber\\
 &&+f_{IJ}\left[\brc{2i\der_4\brkt{\Sgm^IV^J}
 -\frac{i}{2}\derE V^I\der_4V^J+\frac{i}{2}\derE\der_4 V^IV^J}D^\alp\cY_{T\alp} \right.\nonumber\\
 &&\hspace{12mm}
 +2i\Sgm^IV^J\der_4 D^\alp\cY_{T\alp}+\Hc\bigg] \nonumber\\
 &&-\frac{L_{\rm V}}{2R_{\rm E}^-}\Re\brkt{\der_4 D^\alp\Ups_{\rmT\alp}}
 +\frac{2\cX_\rmT L_{\rm V}}{R_{\rm E}^{-2}}\Re\frac{\bar{S}_{\rm E}\der_4\SE}{\SE^2} \nonumber\\
 &&+\frac{f_{IJ}\cX_\rmT}{R_{\rm E}^-}\bigg[
 2i\der_4\bar{S}_{\rm E}\bar{\cO}_{\rm E}V^I\brkt{\derE V^J-2\Sgm^J}
 +4i\der_4\bar{\Sgm}^I\brkt{\derE V^J-\Sgm^J}
 +2iR_{\rm E}^+\der_4\brkt{\Sgm^I\Sgm^J} \nonumber\\
 &&\hspace{18mm}\left.
 +4\brkt{\Im\frac{\bar{S}_{\rm E}\der_4\SE}{\SE^2}}\Sgm^I\Sgm^J+\Hc\right] \nonumber\\
 &&-\der_4\left[\brkt{\frac{\cV_\rmT}{\cX_\rmT R_{\rm E}^-}}^{1/2}L_{\rm H}R_{\rm E}^+
 +\brkt{\frac{if_{IJ}}{2\SE}\der_4V^IV^JD^\alp\cY_{\rmT\alp}+\Hc} 
 +\frac{\cX_\rmT L_{\rm V}}{R_{\rm E}^{-2}}R_{\rm E}^+
 \right.\nonumber\\
 &&\hspace{10mm}\left.
 +\frac{f_{IJ}\cX_\rmT}{R_{\rm E}^-}\brc{\frac{2i}{\bar{S}_{\rm E}}\der_4V^I\brkt{\derE V^J-2\Sgm^J}
 +2i\frac{\SE}{\bar{S}_{\rm E}}\Sgm^I\Sgm^J+\Hc}\right] \nonumber\\
 &&-\der_5\left[-\brkt{\frac{\cV_\rmT}{\cX_\rmT R_{\rm E}^-}}^{1/2}L_{\rm H}\abs{\SE}^2
 -\brkt{\frac{if_{IJ}\SE}{2}\der_4 V^IV^JD^\alp\cY_{\rmT\alp}+\Hc}
 -\frac{\cX_\rmT L_{\rm V}}{R_{\rm E}^{-2}}\abs{\SE}^2 \right.\nonumber\\
 &&\hspace{10mm}\left.
 +\frac{f_{IJ}\cX_\rmT}{R_{\rm E}^-}
 \brc{-2i\bar{S}_{\rm E}\der_4V^I\brkt{\derE V^J-2\Sgm^J}
 -2i\abs{\SE}^2\Sgm^I\Sgm^J+\Hc}\right] \nonumber\\
 &&+(\mbox{brane terms}), 
 \label{EOM:U4}
\eea
and the equation of motion for $U^5$ as 
\bea
 0 \eql -2\Im\Sgm_\rmT\der_5\brc{\brkt{\frac{R_{\rm E}^-}{\cV_\rmT\cX_\rmT}}^{1/2}L_{\rm H}}
 +2\brkt{\frac{\cV_\rmT}{\cX_\rmT R_{\rm E}^-}}^{1/2}L_{\rm H}\Re\frac{\bar{S}_{\rm E}\der_5\SE}{\SE^2} 
 \nonumber\\
 &&-\frac{1}{2}\brkt{\frac{\cV_\rmT R_{\rm E}^-}{\cX_{\rm T}^3}}^{1/2}L_{\rm H}
 \Re\brkt{\der_5 D^\alp\Ups_{\rmT\alp}} \nonumber\\
 &&+4\brkt{\frac{\cV_\rmT R_{\rm E}^-}{\cX_\rmT}}^{1/2}
 \Im\brkt{H_{\rm odd}^\dagger\tl{d}e^{V}\der_5 H_{\rm odd}
 +H_{\rm even}^\dagger\tl{d}e^{-V}\der_5 H_{\rm even}} \nonumber\\
 &&+f_{IJ}\left[\brc{2i\der_5\brkt{\Sgm^IV^J}
 -\frac{i}{2}\derE V^I\der_5V^J+\frac{i}{2}\derE\der_5 V^IV^J}D^\alp\cY_{T\alp} \right.\nonumber\\
 &&\hspace{12mm}
 +2i\Sgm^IV^J\der_5 D^\alp\cY_{T\alp}+\Hc\bigg] \nonumber\\
 &&-\frac{L_{\rm V}}{2R_{\rm E}^-}\Re\brkt{\der_5 D^\alp\Ups_{\rmT\alp}}
 +\frac{2\cX_\rmT L_{\rm V}}{R_{\rm E}^{-2}}\Re\frac{\bar{S}_{\rm E}\der_5\SE}{\SE^2} \nonumber\\
 &&+\frac{f_{IJ}\cX_\rmT}{R_{\rm E}^-}\bigg[
 2i\der_5\bar{S}_{\rm E}\bar{\cO}_{\rm E}V^I\brkt{\derE V^J-2\Sgm^J}
 +4i\der_5\bar{\Sgm}^I\brkt{\derE V^J-\Sgm^J}
 +2iR_{\rm E}^+\der_5\brkt{\Sgm^I\Sgm^J} \nonumber\\
 &&\hspace{18mm}\left.
 +4\brkt{\Im\frac{\bar{S}_{\rm E}\der_5\SE}{\SE^2}}\Sgm^I\Sgm^J+\Hc\right] \nonumber\\
 &&-\der_4\left[\brkt{\frac{\cV_\rmT}{\cX_\rmT R_{\rm E}^-}}^{1/2}\frac{L_{\rm H}}{\abs{\SE}^2}
 +\brkt{\frac{if_{IJ}}{2\SE}\der_5V^IV^JD^\alp\cY_{\rmT\alp}+\Hc} 
 +\frac{\cX_\rmT L_{\rm V}}{R_{\rm E}^{-2}\abs{\SE}^2}
 \right.\nonumber\\
 &&\hspace{10mm}\left.
 +\frac{f_{IJ}\cX_\rmT}{R_{\rm E}^-}\brc{\frac{2i}{\bar{S}_{\rm E}}\der_5V^I\brkt{\derE V^J-2\Sgm^J}
 -\frac{2i}{\abs{\SE}^2}\Sgm^I\Sgm^J+\Hc}\right] \nonumber\\
 &&-\der_5\left[-\brkt{\frac{\cV_\rmT}{\cX_\rmT R_{\rm E}^-}}^{1/2}L_{\rm H}R_{\rm E}^+
 -\brkt{\frac{if_{IJ}\SE}{2}\der_5 V^IV^J D^\alp\cY_{T\alp}+\Hc}
 -\frac{\cX_\rmT L_{\rm V}}{R_{\rm E}^{-2}}R_{\rm E}^+ \right.\nonumber\\
 &&\hspace{10mm}\left.
 +\frac{f_{IJ}\cX_\rmT}{R_{\rm E}^-}\brc{-2i\bar{S}_{\rm E}\der_5V^I\brkt{\derE V^J-2\Sgm^J}
 +2i\frac{\bar{S}_{\rm E}}{\SE}\Sgm^I\Sgm^J+\Hc}\right] \nonumber\\
 &&+(\mbox{brane terms}). 
 \label{EOM:U5}
\eea
The ellipses denote the contributions from the brane terms.

\section{Total flux quantization} \label{FQ:sphere}
Here we give a comment on the quantization of the total flux. 
For this purpose, we begin with solving the Maxwell equation in the bulk, 
\be
 \der_P\brkt{\sqrt{\det G_{MN}}\sgm F^{PQ}} = 0. 
 \label{Maxwell_eq}
\ee
Since we assume that the dilaton~$\sgm$ has a constant background, 
and the background metric~$G_{MN}$ is given by 
\be
 G_{MN}dx^M dx^N = \eta_{\mu\nu}dx^\mu dx^\nu+e^{(2)}\brc{(dx^4)^2+(dx^5)^2}, 
\ee
the above equation becomes
\be
 \sgm\der_m\brkt{\frac{F_{45}}{e^{(2)}}} = 0. 
\ee
Namely, the background field strength~$F_{45}$ is proportional to $e^{(2)}$. 
By solving the Einstein equation, we find that $e^{(2)}$ in (\ref{bgd:solution:15}) is a solution~\cite{Redi:2004tm}. 
Thus, we obtain 
\be
 F_{z\bar{z}} = -2iF_{45} = \frac{ic_F\abs{w'}^2}{\brkt{1+\abs{w}^2}^2}, 
 \label{F:bulk}
\ee
where $c_F$ is an integration constant. 
Since (\ref{Maxwell_eq}) does not include the brane contributions, 
(\ref{F:bulk}) does not have the localized flux terms, 
in contrast to (\ref{expr:Fzbz}). 
Up to the brane-localized terms, (\ref{F:bulk}) can be solved as
\be
 A_z %= \frac{ic_F}{4}\brkt{\frac{w''}{w'}-\frac{2\bar{w}w'}{1+\abs{w}^2}} 
 = \frac{ic_F}{4}\der_z\ln\frac{\abs{w'}^2}{\brkt{1+\abs{w}^2}^2}. 
 \label{assump:A_z}
\ee
In fact, this is the solution including the brane contributions. 
Hence, (\ref{F:bulk}) is modified as 
\be
 F_{z\bar{z}} = \der_z A_{\bar{z}}-\der_{\bar{z}}A_z 
 = \frac{ic_F\abs{w'}^2}{\brkt{1+\abs{w}^2}^2}+i\pi c_F\sum_k\alp_k\dlt^{(2)}(z-z_k). 
\ee
Thus the total flux is calculated as
\be
 \cB = -i\int d^2z\;cF_{z\bar{z}} = \pi cc_F(2-2g), 
 \label{rel:cB-c_F}
\ee
where $g$ is the genus of the compact space. 
We have used (\ref{volume}). 
The coefficient~$c_F$ (or the total flux~$\cB$) will be quantized by requiring the single-valuedness of the fields.

\subsection{Sphere compactification ($\bdm{g=0}$)}
The sphere is covered by the following two coordinate patches. 
\begin{description}
\item[Patch I:] $z$, which covers the whole points except for the infinity. 
 
\item[Patch II:] $\tl{z}\equiv -1/z$, which covers the whole points except for the origin~$z=0$. 
\end{description}
The gauge 1-form~$A$ is expressed in the patch I as~\footnote{
We omit the 4D components of $A$ since they do not have non-vanishing background. 
}
\be
 A^{\rm (I)} = A_zdz+A_{\bar{z}}d\bar{z} 
% = \frac{i}{4c}\brkt{\frac{w''}{w'}-\frac{2\bar{w}w'}{1+\abs{w}^2}} 
 = \frac{i\cB}{8\pi c}dz\der_z\ln\frac{\abs{dw/dz}^2}{\brkt{1+\abs{w}^2}^2}+\Hc, 
\ee
where (\ref{rel:cB-c_F}) has been used, 
while it is expressed in the patch II as
\be
 A^{\rm (II)} = \frac{i\cB}{8\pi c}d\tl{z}\der_{\tl{z}}\ln\frac{\abs{dw/d\tl{z}}^2}{\brkt{1+\abs{w}^2}^2}+\Hc. 
\ee
Since 
\be
 \frac{dw}{dz} = \frac{d\tl{z}}{dz}\frac{dw}{d\tl{z}} = \tl{z}^2\frac{dw}{d\tl{z}}, 
\ee
$A^{\rm (I)}$ can be rewritten as
\bea
 A^{\rm (I)} \eql \frac{i\cB}{8\pi c}d\tl{z}\der_{\tl{z}}
 \ln\frac{\abs{\tl{z}}^4\abs{dw/d\tl{z}}^2}{\brkt{1+\abs{w}^2}^2}+\Hc
 = A^{\rm (II)}+\frac{i\cB}{4\pi c}\brkt{\frac{d\tl{z}}{\tl{z}}-\frac{d\bar{\tl{z}}}{\bar{\tl{z}}}} \nonumber\\
 \eql A^{\rm (II)}+\frac{i\cB}{4\pi c}d\ln\brkt{\frac{\tl{z}}{\bar{\tl{z}}}}
 = A^{\rm (II)}-\frac{\cB}{2\pi c}d\vth,  
\eea
where $\vth\equiv\arg\tl{z}$. 
Thus, from the patch II to the patch I, the gauge transformation:\footnote{
This can be read off by substituting $\Lmd=\frac{i}{2}\lmd$ in (\ref{gauge_trf}). 
} 
\be
 A_M \to A_M+\der_M\lmd, \;\;\;\;\;
 \phi_2^2 \to e^{ic\lmd}\phi_2^2, 
\ee
with $\lmd=-\cB\vth/(2\pi c)$ has to be performed. 
Therefore, from the single-valuedness of $\phi_2^2$, we obtain the flux quantization condition, 
\be
 \cB = 2n\pi. \;\;\;\;\; (n \in {\mathbb Z})
\ee

\subsection{Torus compactification ($\bdm{g=1}$)}
In this case, the total flux vanishes, and the coefficient~$c_F$ is not quantized 
by the requirement that the fields are single-valued. 
In fact, (\ref{assump:A_z}) is invariant under the translations~$z\to z+1\to z+1+\tau\to z+\tau\to z$, 
in contrast to the non-BPS background in the absence of the branes~\cite{Cremades:2004wa}:  
\be
 A_z = -\frac{i\cB\bar{z}}{2c\Im\tau}. 
\ee
In order to fix $c_F$, we have to take into account the other equations of motion.

\ignore{
\subsection{Sphere compactification}
We can choose the origin of the coordinates so that no brane is located at the infinity.  
Then, consider a closed path~$\cC$ that encircles all the brane positions. 
After going around the origin along $\cC$, 
the phase of $\tl{\phi}_2^2$ changes by
\bea
 c\oint_{\cC}\brkt{dz\;\der_z\lmd+d\bar{z}\;\der_{\bar{z}}\lmd} \eql 
 -i\int_{\cS}d^2z\;cF_{z\bar{z}}, 
\eea
where $\cS$ is the region encircled by $\cC$. 
When the radius of $\cC$ is taken large enough, the right-hand side becomes equal to the total flux~$\cB$. 
We should note that all the fields must be single-valued on $\cC$ 
because the infinity is not a singular point. 
Thus, from the requirement for the single-valuedness of $\tl{\phi}_2^2$, we obtain 
the flux quantization condition, 
\be
 \cB = 2n\pi. \;\;\;\;\;
 (n \in {\mathbb Z}) \label{FQ:cond}
\ee
Needless to say, this condition does not depend on the choice of the coordinates. 
\subsection{Torus compactification}
When the compact space is a torus, 
the closed path~$\cC$ is chosen as the boundary of the fundamental region of the torus. 
Then we obtain the same quantization condition~(\ref{FQ:cond}). 
}

\section{Weierstrass elliptic functions} \label{def:Welliptic}
The Weierstrass p-, zeta- and sigma-functions are defined as
\bea
 \wp_{\omg_1,\omg_2}(z) \defa \frac{1}{z^2}
 +\sum_{(m,n)\neq (0,0)}\brc{\frac{1}{(z-\Omg_{m,n})^2}-\frac{1}{\Omg_{m,n}^2}}, \nonumber\\
 \zeta_{\omg_1,\omg_2}(z) \defa \frac{1}{z}
 +\sum_{(m,n)\neq (0,0)}\brkt{\frac{1}{z-\Omg_{m,n}}+\frac{1}{\Omg_{m,n}}+\frac{z}{\Omg_{m,n}^2}}, 
 \nonumber\\
 \sgm_{\omg_1,\omg_2}(z) \defa z\prod_{(m,n)\neq (0,0)}
 \brkt{1-\frac{z}{\Omg_{m,n}}}\exp\brkt{\frac{z}{\Omg_{m,n}}+\frac{z^2}{2\Omg_{m,n}^2}}, 
 \label{def:wp}
\eea
where $\Omg_{m,n}\equiv m\omg_1+n\omg_2$, and $m,n\in{\mathbb Z}$. 
Clearly, $\wp(z)$ is doubly periodic. 
\be
 \wp_{\omg_1,\omg_2}(z+\omg_1) = \wp_{\omg_1,\omg_2}(z+\omg_2) = \wp_{\omg_1,\omg_2}(z). 
\ee
It also follows that
\bea
 \zeta_{\omg_1,\omg_2}'(z) \eql -\wp(z), \nonumber\\
 \zeta_{\omg_1,\omg_2}(z) \eql \frac{d\ln\sgm_{\omg_1,\omg_2}}{dz}(z)
 = \frac{\sgm'_{\omg_1,\omg_2}(z)}{\sgm_{\omg_1,\omg_2}(z)},  \nonumber\\
 \wp_{\omg_1,\omg_2}\brkt{z+\frac{\omg_1}{2}} \eql 
 e_1+\frac{(e_1-e_2)(e_1-e_3)}{\wp_{\omg_1,\omg_2}(z)-e_1}, \nonumber\\
 \wp_{\omg_1,\omg_2}\brkt{z+\frac{\omg_2}{2}} \eql 
 e_2+\frac{(e_2-e_1)(e_2-e_3)}{\wp_{\omg_1,\omg_2}(z)-e_2}, 
 \label{formulae:wp}
\eea
where $e_1\equiv\wp_{\omg_1,\omg_2}(\omg_1/2)$, $e_2\equiv\wp_{\omg_1,\omg_2}(\omg_2/2)$, 
and $e_3=-e_1-e_2$.

%%%%%%%%%%%%%%%%%%%%%%%%%%%% References %%%%%%%%%%%%%%%%%%%%%%%%%%%%%%


\begin{thebibliography}{99}
\bibitem{Marcus:1983wb}
  N.~Marcus, A.~Sagnotti and W.~Siegel,
  %``Ten-dimensional Supersymmetric {Yang-Mills} Theory in Terms of Four-dimensional Superfields,''
  Nucl.\ Phys.\ B {\bf 224} (1983) 159. 
%  doi:10.1016/0550-3213(83)90318-8

\bibitem{ArkaniHamed:2001tb}
  N.~Arkani-Hamed, T.~Gregoire and J.~G.~Wacker,
  %``Higher dimensional supersymmetry in 4-D superspace,''
  JHEP {\bf 0203} (2002) 055
%  doi:10.1088/1126-6708/2002/03/055
  [hep-th/0101233]. 

\bibitem{Marti:2001iw}
  D.~Marti and A.~Pomarol,
  %``Supersymmetric theories with compact extra dimensions in N=1 superfields,''
  Phys.\ Rev.\ D {\bf 64} (2001) 105025
%  doi:10.1103/PhysRevD.64.105025
  [hep-th/0106256]. 

\bibitem{Hebecker:2001ke}
  A.~Hebecker,
  %``5-D superYang-Mills theory in 4-D superspace, superfield brane operators, and applications to orbifold GUTs,''
  Nucl.\ Phys.\ B {\bf 632} (2002) 101
%  doi:10.1016/S0550-3213(02)00253-5
  [hep-ph/0112230]. 

\bibitem{Abe:2012ya}
  H.~Abe, T.~Kobayashi, H.~Ohki and K.~Sumita,
  %``Superfield description of 10D SYM theory with magnetized extra dimensions,''
  Nucl.\ Phys.\ B {\bf 863} (2012) 1
%  doi:10.1016/j.nuclphysb.2012.05.012
  [arXiv:1204.5327 [hep-th]]. 

\bibitem{Abe:2015jqa}
  H.~Abe, T.~Horie and K.~Sumita,
  %``Superfield description of (4+2n)-dimensional SYM theories and their mixtures on magnetized tori,''
  Nucl.\ Phys.\ B {\bf 900} (2015) 331
%  doi:10.1016/j.nuclphysb.2015.09.014
  [arXiv:1507.02425 [hep-th]]. 

\bibitem{Linch:2002wg}
  W.~D.~Linch, III, M.~A.~Luty and J.~Phillips,
  %``Five-dimensional supergravity in N=1 superspace,''
  Phys.\ Rev.\ D {\bf 68} (2003) 025008
%  doi:10.1103/PhysRevD.68.025008
  [hep-th/0209060]. 

\bibitem{Paccetti:2004ri}
  F.~Paccetti Correia, M.~G.~Schmidt and Z.~Tavartkiladze,
  %``Superfield approach to 5D conformal SUGRA and the radion,''
  Nucl.\ Phys.\ B {\bf 709} (2005) 141
%  doi:10.1016/j.nuclphysb.2004.12.005
  [hep-th/0408138]. 

\bibitem{Abe:2004ar}
  H.~Abe and Y.~Sakamura,
  %``Superfield description of 5-D supergravity on general warped geometry,''
  JHEP {\bf 0410} (2004) 013
%  doi:10.1088/1126-6708/2004/10/013
  [hep-th/0408224]. 

\bibitem{Bergshoeff:1985mz}
  E.~Bergshoeff, E.~Sezgin and A.~Van Proeyen,
  %``Superconformal Tensor Calculus and Matter Couplings in Six-dimensions,''
  Nucl.\ Phys.\ B {\bf 264} (1986) 653
   Erratum: [Nucl.\ Phys.\ B {\bf 598} (2001) 667]. 
%  doi:10.1016/0550-3213(86)90503-1

\bibitem{Coomans:2011ih}
  F.~Coomans and A.~Van Proeyen,
  %``Off-shell N=(1,0), D=6 supergravity from superconformal methods,''
  JHEP {\bf 1102} (2011) 049
   Erratum: [JHEP {\bf 1201} (2012) 119]
%  doi:10.1007/JHEP02(2011)049, 10.1007/JHEP01(2012)119
  [arXiv:1101.2403 [hep-th]]. 

\bibitem{Abe:2015bqa}
  H.~Abe, Y.~Sakamura and Y.~Yamada,
  %``N =1 superfield description of vector-tensor couplings in six dimensions,''
  JHEP {\bf 1504} (2015) 035
%  doi:10.1007/JHEP04(2015)035
  [arXiv:1501.07642 [hep-th]]. 

\bibitem{Abe:2015yya}
  H.~Abe, Y.~Sakamura and Y.~Yamada,
  %``$ \mathcal{N}=1 $ superfield description of six-dimensional supergravity,''
  JHEP {\bf 1510} (2015) 181
%  doi:10.1007/JHEP10(2015)181
  [arXiv:1507.08435 [hep-th]]. 

\bibitem{Abe:2017pvw}
  H.~Abe, S.~Aoki and Y.~Sakamura,
  %``Full diffeomorphism and Lorentz invariance in 4D $ \mathcal{N}=1 $ superfield description of 6D SUGRA,''
  JHEP {\bf 1711} (2017) 146
 % doi:10.1007/JHEP11(2017)146
  [arXiv:1708.09106 [hep-th]]. 

\bibitem{Abe:2006eg}
  H.~Abe and Y.~Sakamura,
  %``Roles of Z(2)-odd N=1 multiplets in off-shell dimensional reduction of 5D supergravity,''
  Phys.\ Rev.\ D {\bf 75} (2007) 025018
%  doi:10.1103/PhysRevD.75.025018
  [hep-th/0610234]. 

\bibitem{Abe:2008an}
  H.~Abe and Y.~Sakamura,
  %``Flavor structure with multi moduli in 5D supergravity,''
  Phys.\ Rev.\ D {\bf 79} (2009) 045005
%  doi:10.1103/PhysRevD.79.045005
  [arXiv:0807.3725 [hep-th]]. 

\bibitem{Abe:2011rg}
  H.~Abe, H.~Otsuka, Y.~Sakamura and Y.~Yamada,
  %``SUSY Flavor Structure of Generic 5D Supergravity Models,''
  Eur.\ Phys.\ J.\ C {\bf 72} (2012) 2018
%  doi:10.1140/epjc/s10052-012-2018-x
  [arXiv:1111.3721 [hep-ph]]. 

\bibitem{Sakamura:2013cqd}
  Y.~Sakamura,
  %``One-loop Kaehler potential in 5D gauged supergravity with generic prepotential,''
  Nucl.\ Phys.\ B {\bf 873} (2013) 165
   Erratum: [Nucl.\ Phys.\ B {\bf 873} (2013) 728]
%  doi:10.1016/j.nuclphysb.2013.05.006, 10.1016/j.nuclphysb.2013.04.013
  [arXiv:1302.7244 [hep-th]]. 

\bibitem{Sakamura:2013rba}
  Y.~Sakamura and Y.~Yamada,
  %``Impacts of non-geometric moduli on effective theory of 5D supergravity,''
  JHEP {\bf 1311} (2013) 090
   Erratum: [JHEP {\bf 1401} (2014) 181]
%  doi:10.1007/JHEP11(2013)090, 10.1007/JHEP01(2014)181
  [arXiv:1307.5585 [hep-th]]. 

\bibitem{Sakamura:2014aja}
  Y.~Sakamura and Y.~Yamada,
  %``Natural realization of a large extra dimension in 5D supersymmetric theory,''
  PTEP {\bf 2014} (2014) no.9,  093B02
%  doi:10.1093/ptep/ptu114
  [arXiv:1401.1921 [hep-ph]]. 

\bibitem{Carroll:2003db}
  S.~M.~Carroll and M.~M.~Guica,
  %``Sidestepping the cosmological constant with football shaped extra dimensions,''
  hep-th/0302067. 

\bibitem{Navarro:2003vw}
  I.~Navarro,
  %``Codimension two compactifications and the cosmological constant problem,''
  JCAP {\bf 0309} (2003) 004
%  doi:10.1088/1475-7516/2003/09/004
  [hep-th/0302129]. 

\bibitem{Navarro:2003bf}
  I.~Navarro,
  %``Spheres, deficit angles and the cosmological constant,''
  Class.\ Quant.\ Grav.\  {\bf 20} (2003) 3603
%  doi:10.1088/0264-9381/20/16/306
  [hep-th/0305014]. 

\bibitem{Gibbons:2003di}
  G.~W.~Gibbons, R.~Gueven and C.~N.~Pope,
  %``3-branes and uniqueness of the Salam-Sezgin vacuum,''
  Phys.\ Lett.\ B {\bf 595} (2004) 498
%  doi:10.1016/j.physletb.2004.06.048
  [hep-th/0307238]. 

\bibitem{Lee:2005az}
  H.~M.~Lee and C.~Ludeling,
  %``The General warped solution with conical branes in six-dimensional supergravity,''
  JHEP {\bf 0601} (2006) 062
%  doi:10.1088/1126-6708/2006/01/062
  [hep-th/0510026]. 

\bibitem{Wess:1992cp}
  J.~Wess and J.~Bagger,
  %``Supersymmetry and supergravity,''
  Princeton, USA: Univ. Pr. (1992) 259 p. 

\bibitem{RandjbarDaemi:1985wc}
  S.~Randjbar-Daemi, A.~Salam, E.~Sezgin and J.~A.~Strathdee,
  %``An Anomaly Free Model in Six-Dimensions,''
  Phys.\ Lett.\  {\bf 151B} (1985) 351.
%  doi:10.1016/0370-2693(85)91653-3

\bibitem{Green:1984bx}
  M.~B.~Green, J.~H.~Schwarz and P.~C.~West,
  %``Anomaly Free Chiral Theories in Six-Dimensions,''
  Nucl.\ Phys.\ B {\bf 254} (1985) 327.
%  doi:10.1016/0550-3213(85)90222-6

\bibitem{Kumar:2010ru}
  V.~Kumar, D.~R.~Morrison and W.~Taylor,
  %``Global aspects of the space of 6D N = 1 supergravities,''
  JHEP {\bf 1011} (2010) 118
%  doi:10.1007/JHEP11(2010)118
  [arXiv:1008.1062 [hep-th]]. 

\bibitem{Sakamura:2011df}
  Y.~Sakamura,
  %``Direct relation of linearized supergravity to superconformal formulation,''
  JHEP {\bf 1112} (2011) 008
%  doi:10.1007/JHEP12(2011)008
  [arXiv:1107.4247 [hep-th]]. 

\bibitem{Horvathy:1998pe}
  P.~A.~Horvathy and J.~C.~Yera,
  %``Vortex solutions of the Liouville equation,''
  Lett.\ Math.\ Phys.\  {\bf 46} (1998) 111
%  doi:10.1023/A:1007500510018
  [hep-th/9805161]. 

\bibitem{Redi:2004tm}
  M.~Redi,
  %``Footballs, conical singularities and the Liouville equation,''
  Phys.\ Rev.\ D {\bf 71} (2005) 044006
%  doi:10.1103/PhysRevD.71.044006
  [hep-th/0412189]. 

\bibitem{Akerblom:2009ev}
  N.~Akerblom, G.~Cornelissen, G.~Stavenga and J.~W.~van Holten,
  %``Nonrelativistic Chern-Simons Vortices on the Torus,''
  J.\ Math.\ Phys.\  {\bf 52} (2011) 072901
%  doi:10.1063/1.3610643
  [arXiv:0912.0718 [hep-th]]. 

\bibitem{Akerblom:2010xb}
  N.~Akerblom and G.~Cornelissen,
  %``A Compact Codimension Two Braneworld with Precisely One Brane,''
  Phys.\ Rev.\ D {\bf 81} (2010) 124025
%  doi:10.1103/PhysRevD.81.124025
  [arXiv:1004.1807 [hep-th]]. 

\bibitem{Salam:1984cj}
  A.~Salam and E.~Sezgin,
  %``Chiral Compactification on Minkowski x S**2 of N=2 Einstein-Maxwell Supergravity in Six-Dimensions,''
  Phys.\ Lett.\ B {\bf 147} (1984) 47
   [Phys.\ Lett.\  {\bf 147B} (1984) 47].
%  doi:10.1016/0370-2693(84)90589-6

\bibitem{Randall:1999ee}
  L.~Randall and R.~Sundrum,
  %``A Large mass hierarchy from a small extra dimension,''
  Phys.\ Rev.\ Lett.\  {\bf 83} (1999) 3370
%  doi:10.1103/PhysRevLett.83.3370
  [hep-ph/9905221]. 

\bibitem{Olesen:1991df}
  P.~Olesen,
  %``Vacuum structure of the electroweak theory in high magnetic fields,''
  Phys.\ Lett.\ B {\bf 268} (1991) 389.
%  doi:10.1016/0370-2693(91)91595-M

\bibitem{Olesen:1991dg}
  P.~Olesen,
  %``Soliton condensation in some selfdual Chern-Simons theories,''
  Phys.\ Lett.\ B {\bf 265} (1991) 361
   Erratum: [Phys.\ Lett.\ B {\bf 267} (1991) 541]. 
%  doi:10.1016/0370-2693(91)90066-Y

\bibitem{Cremades:2004wa}
  D.~Cremades, L.~E.~Ibanez and F.~Marchesano,
  %``Computing Yukawa couplings from magnetized extra dimensions,''
  JHEP {\bf 0405} (2004) 079
%  doi:10.1088/1126-6708/2004/05/079
  [hep-th/0404229]. 
\end{thebibliography}
\end{document}